\documentclass{aa}  

\usepackage{graphicx}
\usepackage{txfonts}
\usepackage{xcolor} 
\usepackage{ulem} 

\begin{document}

   \title{Giant planet formation at the pressure maxima of protoplanetary disks}

\subtitle{II. A hybrid accretion scenario}

   \author{Octavio Miguel Guilera 
          \inst{1,2,3}
          \and
          Zsolt S\'andor
          \inst{4,5}
          \and 
          Mar\'{\i}a Paula Ronco 
          \inst{2,3}
          \and 
          Julia Venturini
          \inst{6}
          \and
          Marcelo Miguel Miller Bertolami
          \inst{1,7}
          }
          
   \institute{Grupo de Astrof\'{\i}sica Planteria, Instituto de Astrof\'{\i}sica de La Plata, CCT La Plata-CONICET-UNLP, Paseo del Bosque S/N (1900), La Plata, Argentina.\\
     \email{oguilera@fcaglp.unlp.edu.ar}
     \and
     Instituto de Astrof\'{\i}sica, Pontificia Universidad Cat\'olica de Chile, Santiago, Chile.\\
     \email{mronco@astro.puc.cl}
     \and
     N\'ucleo Milenio Formaci\'on Planetaria (NPF), Pontificia Universidad Cat\'olica, Chile.
     \and
     Department of Astronomy,
     E\"otv\"os Lor\'and University,
     H-1117 Budapest, P\'azm\'any P\'eter s\'et\'any 1/A, Hungary.\\
     \email{Zs.Sandor@astro.elte.hu}
     \and
     Konkoly Observatory,
     Research Centre for Astronomy and Earth Sciences,
     H-1121 Budapest, Konkoly Thege Miklós út 15-17, Hungary.
     \and
     International Space Science Institute, Hallerstrasse 6, 3012, Bern, Switzerland.\\
     \email{julia.venturini@issibern.ch}
     \and
     Facultad de Ciencias Astron\'omicas y Geof\'{\i}sicas, Universidad Nacional de La Plata, Paseo del Bosque S/N (1900), La Plata, Argentina.\\
     \email{mmiller@fcaglp.unlp.edu.ar}
   }
   \date{Received ...; accepted ...}
 
   
   \abstract
       {Recent high-resolution observations of protoplanetary disks have revealed ring-like structures that can be associated to pressure maxima. Pressure maxima are known to be dust collectors and planet migration traps. The great majority of planet formation works are based either on the pebble accretion model or on the planetesimal accretion model. However, recent studies proposed the possible formation of Jupiter by the hybrid accretion of pebbles and planetesimals.}
       {We aim to study the full process of planet formation consisting of dust evolution, planetesimal formation and planet growth at a pressure maximum in a protoplanetary disk.}
       {We compute, through numerical simulations, the gas and dust evolution in a protoplanetary disk, including dust growth, fragmentation, radial drift and particle accumulation at a pressure maximum. The pressure maximum appears due to an assumed viscosity transition at the water ice-line. We also consider the formation of planetesimals by streaming instability and the formation of a moon-size embryo that grows into a giant planet by the hybrid accretion of pebbles and planetesimals, all within the pressure maximum.}
       {We find that the pressure maximum is an efficient collector of dust drifting inwards. The condition of planetesimal formation by streaming instability is fulfilled due to the large amount of dust accumulated at the pressure bump. Then, a massive core is quickly formed (in $\sim 10^4$ yr) by the accretion of pebbles. After the pebble isolation mass is reached, the growth of the core slowly continues by the accretion of planetesimals. The energy released by planetesimal accretion delays the onset of runaway gas accretion, allowing a gas giant to form after $\sim$1 Myr of disk evolution. The pressure maximum also acts as a migration trap.} 
       {Pressure maxima generated by a viscosity transition at the water ice-line are preferential locations for dust traps, planetesimal formation by streaming instability and planet migration traps. All these conditions allow the fast formation of a giant planet by the hybrid accretion of pebbles and planetesimals.}
       

   \maketitle
%

\section{Introduction}

The study of planetary formation is one of the main topics in modern astronomy, specially since the discovery of the first exoplanet in orbit around a solar-type star \citet{Mayor1995}. Despite of the fact that our knowledge on planet formation significantly increased in the last decades, some fundamental issues remain open. Up to date, there are two main planet formation theories: the core accretion mechanism \citep{Safronov1969, Mizuno1980, bodenheimer1986} and the gravitational instability model \citep{Boss1997}. Core accretion is currently the most accepted paradigm, mainly because it naturally explains the diversity of observed planets \citep[see][and reference therein]{Benz2014,Venturini2020a}. In the core accretion mechanism there are two main approaches: the classical planetesimal accretion model \citep{P96, Alibert2005, Fortier07, Fortier2009, Guilera2010, Ormel2012, Alibert13, Guilera2014, Venturini+2015, Venturini+2016, Mordasini2018}, and the  more recent pebble accretion model \citep{Ormel&Klahr2010, Lambrechts&Johansen2012, Lambrechts+2014, JohansenLambrechts2017, Ndugu18, Lambrechts19}.
 
In most planet formation simulations the most important physical properties of disks, such as the surface densities of gas and solids and the temperature are characterized by a monotonically decreasing power law profile.
However, recent observations of protoplanetary disks by ALMA (Atacama Large Millimeter Array) reveal that disks are rich in substructures, such as rings \citep{DSHARP1}. Rings are dust concentrations that arise naturally by the presence of pressure bumps \citep{Dullemond2018}. 

From a theoretical point of view, there are some crucial obstacles to grow planets from micrometer-size dust grains. The first is the so-called meter-size barrier, which refers to the difficulty to grow particles of meter size due to the combined effect of fast radial drift and destructive collisions \citep[see e.g,][]{blum_wurm2008}. Another important problem is the rapid inward planet migration, that makes it hard for planets to remain at a few astronomical units from the star \citep{IdaLin2008,Miguel2011a}.
One possible solution is the existence of preferential places for dust and planet growth, known, respectively, as dust and planet traps. In dust traps the drag force on dust grains vanishes, while in planet traps the torque exerted by the disk onto the planet is zero and migration is therefore halted. Particularly interesting is the case when a planet trap is generated by a density maximum, since in general (through the equation of state that links density and pressure) a pressure maximum is also developed. Pressure maxima act as dust traps, collecting grains, and helping their further growth to planetesimal, or even to embryo sizes via either coagulation \citep{brauer_etal2008b}, or drag-induced instabilities \citep{YoudinShu2002ApJ,Johansenetal2009ApJ}. If due to the subsequent accretion processes massive embryos form in a pressure maximum, they might be locked in the nearby developed planet trap. The combined effect of a planet trap and pressure maximum in promoting the formation of larger bodies, with sizes ranging between Mars and Jupiter, has been investigated by \citet{Lyraetal2008,lyra_etal2009,sandor_etal2011, Regalyetal2013MNRAS}, and \citet{Guilera&Sandor2017}. Pressure traps are naturally expected at locations in the disk where a transition in the accretion flow occurs. Such transition can develop at the boundary between an accretionally active region, in which the magneto-rotational instability (MRI) might be dominant, and an accretionally less active region (usually called dead zone). A similar phenomenon occurs at the ice-line, where below certain temperature (usually of about 170 K for typical disk conditions) water vapor is condensed in the form of fluffy ice-flakes that due to their high surface-to-volume ratio efficiently trap free electrons in the plasma reducing its conductivity \citep{Sano2000,IlgnerNelson2006,Okuzumi2009,Dzyurkevich2013}. Based on these results, \citet{kretke_lin2007} and \citet{brauer_etal2008b} consider the ice-line the location in the disk where the meter-size barrier problem could be overcome. The reduced conductivity leads to a steep decrease of the alpha viscosity parameter as has been shown by \citet{Lyra2015}
in magneto-hydrodynamical simulations (MHD), if the MRI is assumed to be the main reason for gas accretion. Due to the reduced turbulent viscosity, as the disk is tending to adapt to a new steady state, a maximum in surface density, and through the equation of state, a pressure maximum develops. Thus, the water ice-line has been considered as an ideal place in where the meter size barrier could be overtaken.

To study in detail the formation of a planet within a pressure maximum by concurrent solid and gaseous accretion, one must combine a model of pebble and planetesimal growth and evolution and their accretion onto a planetary embryo, coupled with the cooling of the gaseous envelope and gas supply from the disk. The accretion of solids by a protoplanet is typically studied under the simplifying assumption  of a unique dominant solid's size: either cm-pebbles or km-planetesimals. Recently, \citet{alibert2018} and \citet{Venturini-Helled2020} proposed and studied the formation of Jupiter by a hybrid accretion of pebbles and planetesimals. In this scenario, the core is formed fast by pebble accretion, and after the planet reaches the pebble isolation mass, the heat released by the accretion of planetesimals delays the onset of the gaseous runaway.

This hybrid accretion model could explain some measurements of isotopes in iron meteorites that provided evidence for the existence of two reservoirs of small bodies in the early Solar System, which remained separated from $\sim1$~Myr to $\sim3$~Myr after the CAIs (calcium-aluminium-inclusions) formation \citep{Kruijer2017}. It could also account for the bulk and atmospheric enrichment of Jupiter reported by recent structure models by \citet{Wahl2017} \citep{Venturini-Helled2020}. However, these works adopted some important simplifications: the massive core of the planet was already formed, only the accretion of planetesimals and gas was computed. Also, as it is usually assumed in core accretion simulations for simplification, the planetesimals were placed initially in the disk without accounting for their formation.\\ 

The main goal of this work is to extend our previous work \citep{Guilera&Sandor2017}, where we studied the formation of giant planets at the pressure maxima generated at the edges of a dead zone, incorporating a model of dust evolution, dust growth, planetesimal formation and a hybrid accretion of pebbles and planetesimals to study the formation of a giant planet at a pressure maximum in the disk. The work is organized in the following way: in Sect. 2 we briefly describe our planet formation model with its new improvements, in Sect. 3 we present our results, in Sect. 4 we discuss our findings, and finally in Sect. 5 we draw our conclusions.


\section{The model}
\label{sec:sec2}
In this work we consider a viscous evolving disk with a transition in the eddy viscosity at the water ice-line, where the presence of ice flakes removes free electrons from the gas, leading to a jump in resistivity, similarly to \citet{Guilera&Sandor2017}. Our planet formation code {\scriptsize PLANETALP} allows us to calculate the formation of planets immersed in a protoplanetary disk that evolves in time. In this work, we also take into account a model of dust growth and evolution, coupled with a model of planetesimals formation by streaming instability. Thus, we are now able to compute the full process of planet formation: from a micron-sized dust population along the disk that evolves in time and forms planetesimals (when the conditions for the streaming instability are fulfilled), to the formation of a moon-size embryo that grows by the simultaneous accretion of pebbles and planetesimals in what is known as a hybrid accretion model \citep{alibert2018}. We also consider the possible migration of the planet by computing the torque onto the planet in the type I migration regime \citep{Tanaka+2002, paardekooper.etal2011, jm2017} and the thermal torque due to the heat released by the planet into the disk as the result of the accretion of solid material \citep{BLlambay2015, Masset2017}. These migration regimes were already included in \citet{Guilera+2019}. In the next sub-sections, we summarize the main characteristics of our model and the improvements incorporated into this work. The rest of the model is described in detail in \citet{Ronco+2017}, \citet{Guilera&Sandor2017}, and \citet{Guilera+2019}.      

\subsection{The gaseous disk}
\label{sec:sec2-0}

As in \citet{Guilera&Sandor2017}, we use a classical 1D radial model (the isothermal version of {\scriptsize PLANETALP} \citep{Ronco+2017}) where the gaseous component of the disk is characterized by the corresponding surface density $\Sigma_\text{g}$. We define an initial disk with mass $0.05~\text{M}_{\odot}$ and initial gas surface density profile given by 
\begin{eqnarray}
  \Sigma_{\text{g}} &=& \Sigma_{\text{g}}^0 \left( \frac{R}{R_c} \right)^{-\gamma} e^{-(R/R_c)^{2-\gamma}}, \label{eq1-sec2-0}
\end{eqnarray}
where $R$ is the radial coordinate, $R_c= 30$~au is the characteristic radius, and $\gamma = 1$ is the surface density exponent \citep{Andrews2010}, while $\Sigma_{\text{g}}^0$ is a normalization parameter that depends on the disk mass. For simplicity, the initial disk domain is defined between 0.1 au and 30 au. The disk evolves as an $\alpha$ accretion disk \citep{Pringle1981}  
\begin{eqnarray}
  \frac{\partial \Sigma_{\text{g}}}{\partial t}= \frac{3}{R}\frac{\partial}{\partial R} \left[ R^{1/2} \frac{\partial}{\partial R} \left( \nu \Sigma_{\text{g}} R^{1/2}  \right) \right], 
\label{eq2-sec2-0}
\end{eqnarray}
where $\nu= \alpha c_s \text{H}_{\text{g}} $ is the kinematic viscosity given by the dimensionless parameter $\alpha$. The sound speed is given by
\begin{eqnarray}
  c_s= \sqrt{ \frac{\gamma_{\text{g}} k_{\text{B}} \text{T}}{\mu m_\text{H}}},
  \label{eq2.1-sec2-0}
\end{eqnarray}  
where $\gamma_{\text{g}}= 7/5$ is the adiabatic constant, $k_{\text{B}}$ is the  Boltzmann-constant, $\mu=2.3$ is the mean molecular weight of molecular gas assuming a typical H-He composition, and $m_\text{H}$ is the proton's mass. The radial temperature profile is given by 
\begin{eqnarray}
  \text{T}= 280 \left( \frac{R}{1~\text{au}} \right)^{-1/2} ~ \text{K}.
  \label{eq2.2-sec2-0}
\end{eqnarray}
The scale height of the disk is given by 
\begin{eqnarray}
\text{H}_\text{g}= c_s/\Omega_{\text{k}},  
   \label{eq2.3-sec2-0}
\end{eqnarray}
$\Omega_{\text{k}}$ being the Keplerian frequency. With the definition of Eq.~(\ref{eq2.3-sec2-0}), the volumetric gas density and the pressure at the disk mid-plane are given by, 
\begin{eqnarray}
\rho_{\text{g}} &=&  \dfrac{\Sigma_{\text{g}}}{\sqrt{2\pi}~\text{H}_\text{g}},\\
\text{P}_{\text{g}} &=& c_s^2 \rho_{\text{g}}. 
 \label{eq2.4-sec2-0}
\end{eqnarray}

As in \citet{Guilera&Sandor2017}, we consider a viscosity transition at the ice-line (at $\sim 3$~au). We note that we use the same functional form to calculate the transition in the $\alpha$-viscosity parameter used in \citet{Guilera&Sandor2017}, but without considering the outer edge of the dead zone. Thus, $\alpha$ takes values of $10^{-3}$ and $10^{-5}$ inside and outside the ice-line, respectively. These are usually applied values for active and dead zones, respectively \citep[e.g.][]{Matsumura2009, Pinilla2016, Yang2018, Ogihara2018}, but their change within some limits does not modify the outcome of this work. In Fig.~1 we illustrate the time evolution of the radial profiles of the gas surface density for our disk. As in \citet{Guilera&Sandor2017}, a density/pressure maximum is formed at the location of the viscosity transition.
\begin{figure}
  \centering
  \includegraphics[width= 0.475\textwidth]{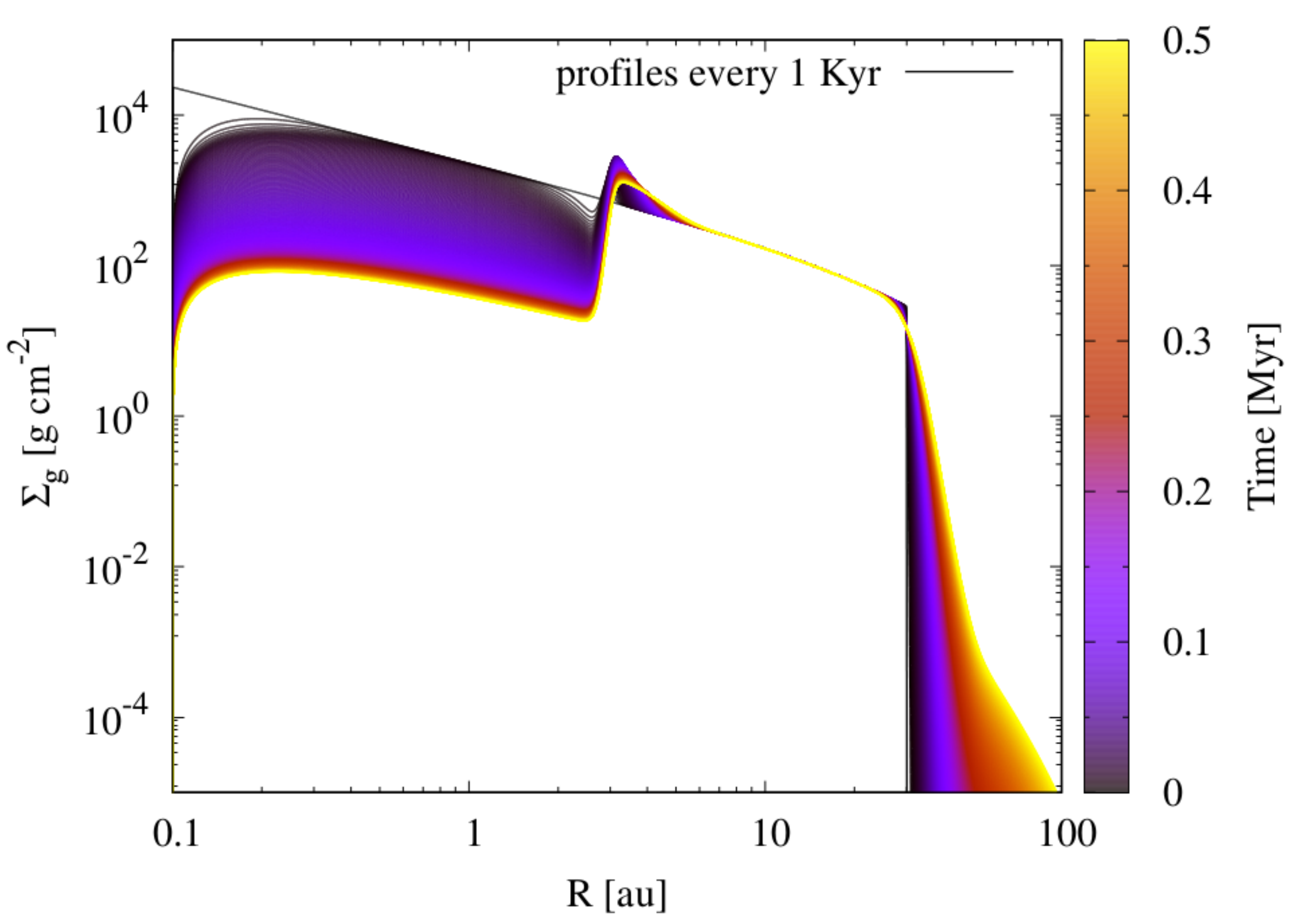}
  \caption{Viscous evolution of the gas surface density radial profiles during the first 0.5~Myr.}
  \label{fig1-sec2-0}
\end{figure}

\subsection{Dust growth}
\label{sec:sec2-1-0}

In order to compute the growth of the dust particles along the disk, we follow the approach derived by \citet{Drazkowska+2016} (hereafter D16) and \citet{Drazkowska+2017}, based on the results of \citet{Birnstiel+2011, Birnstiel+2012}. In this model, the maximum size of the dust particles at each radial bin is limited by the combined effects of the dust coagulation, radial drift and fragmentation. As we consider a region of low-viscosity in the disk, and as it was pointed out by D16, we also include a growth limitation due to the fragmentation induced by differential drift. Thus, the maximum size at a given time is given by
\begin{equation}
  r_{\text{d}}^{\text{max}}(t)= \min(r_{\text{d}}^0 \, \exp(t/\tau_{\text{growth}}), \, r_{\text{drift}}^{\text{max}}, \, r_{\text{frag}}^{\text{max}},
  r_{\text{ddf}}^{\text{max}}),
  \label{eq1-sec2-1-0}
\end{equation}
where $r_{\text{d}}^0= 1~\mu\text{m}$ is the initial dust size. $\tau_{\text{growth}}$ is the collisional growth timescale 
\begin{equation}
  \tau_{\text{growth}}= \dfrac{1}{Z\Omega_{\text{k}}},
  \label{eq2-sec2-1-0}
\end{equation}
$Z= \Sigma_{\text{d}}/\Sigma_{\text{g}}$ being the dust-to-gas ratio. The maximum size of dust particles limited by radial drift is given by 
\begin{equation}
  r_{\text{drift}}^{\text{max}}= f_{\text{d}} \dfrac{2\Sigma_{\text{d}}v_{\text{k}}^2}{\pi \rho'_{\text{d}} c_s^2} \left| \dfrac{d\,\ln P_{\text{g}}}{d\,\ln R} \right|^{-1}, 
  \label{eq3-sec2-1-0}
\end{equation}
where $f_{\text{d}}= 0.55$ \citep{Birnstiel+2012}, $v_{\text{k}}$ is the Keplerian velocity, $\rho'_{\text{d}}$ is the bulk dust density taking values of $3~\text{g}/\text{cm}^3$ and $1~\text{g}/\text{cm}^3$ inside and outside the ice-line, respectively, considering that the dust inside the ice-line is composed fully by silicates and that the dust outside the ice-line is composed by a mixture of silicates and ices \citep{Drazkowska+2017}, $c_s$ is the sound speed in the gas disk, and $P_{\text{g}}$ the gas pressure. Then, the maximum size of dust particles limited by the fragmentation process is set by 
\begin{equation}
  r_{\text{frag}}^{\text{max}}= f_{\text{f}} \dfrac{2\Sigma_{\text{g}}v_{\text{th}}^2}{3\pi \rho'_{\text{d}} \alpha c_s^2}, 
  \label{eq4-sec2-1-0}
\end{equation}
where $f_{\text{f}}= 0.37$ \citep{Birnstiel+2011}, and we use a fragmentation threshold velocity $v_{\text{th}}$ of $1~\text{m}/\text{s}$ for silicate dust and $10~\text{m}/\text{s}$ for dust containing more than 1\% of water ice \citep{Drazkowska+2017}. Eq.~(\ref{eq4-sec2-1-0}) considers that fragmentation is driven by the turbulent velocities. However, if the viscosity in the disk mid-plane becomes very low, fragmentation could be driven by differential drift. Thus, the maximum size of dust particles determined by fragmentation in laminar disks is given by \citep{Birnstiel+2012}
\begin{equation}
  r_{\text{ddf}}^{\text{max}}= \dfrac{4 \Sigma_{\text{g}}v_{\text{th}}v_{\text{k}}}{c_s^2 \pi \rho'_{\text{d}}} \left| \dfrac{d\,\ln P_{\text{g}}}{d\,\ln R} \right|^{-1}. 
  \label{eq5-sec2-1-0}
\end{equation}

\subsection{Dust evolution}
\label{sec:sec2-1}

Due to the fact that the dust is strongly coupled to the dynamics of the gas, the time evolution of the dust solid surface density, $\Sigma_{\text{d}}$, is calculated by using an advection-diffusion equation, 
\begin{equation}
  \frac{\partial}{\partial t} \left(\Sigma_{\text{d}}\right) + \frac{1}{R} \frac{\partial}{\partial R} \left( R \,\overline{v}_{\text{drift}} \,\Sigma_{\text{d}} \right) - \frac{1}{R} \frac{\partial}{\partial R} \left[ R \,D^{*} \,\Sigma_{\text{g}}   \frac{\partial}{\partial R} \left( \frac{\Sigma_{\text{d}}}{\Sigma_{\text{g}}} \right) \right] = \dot{\Sigma}_{\text{d}},  
  \label{eq1-sec2-1}
\end{equation}
where $t$ is the temporal coordinate. $\dot{\Sigma}_{\text{d}}$ represents the sink term due to both planetesimal formation and planet accretion (see next sections). $D^{*}= \nu / ( 1 + \overline{\text{St}}^2 )$ is the dust diffusivity \citep{YoudinLithwick2007}, $\overline{\text{St}}= \pi \rho'_{\text{d}} \overline{r}_{\text{d}} / 2 \Sigma_{\text{g}}$ being the mass weighted mean Stokes number of the dust size distribution (the dust distribution remains always in the Epstein drag regime). In the last expression, $\overline{r}_{\text{d}}$ represents the mass weighted mean radius of the dust size distribution, given by
\begin{equation}
  \overline{r}_{\text{d}}= \dfrac{ \sum_i \epsilon_i r_{\text{d}}^i }{ \sum_i \epsilon_i },  
  \label{eq2-sec2-1}
\end{equation}
$r_{\text{d}}^i$ being the radius of the dust particle of the species $i$, and $\epsilon_i= \rho^{i}_{\text{d}}/\rho_{\text{g}}$ the ratio between the volumetric dust density of the species i and the volumetric gas density given by
\begin{equation}
\epsilon_i= \dfrac{\rho^{i}_{\text{d}}}{\rho_{\text{g}}}= \dfrac{\Sigma^{i}_{\text{d}}}{\Sigma_{\text{g}}} \sqrt{ \dfrac{\alpha + \text{St}^{i}}{\alpha}},
 \label{eq2bis-sec2-1}
\end{equation}
with $\text{St}^{i}= \pi \rho'_{\text{d}} r_{\text{d}}^i / 2 \Sigma_{\text{g}}$ the Stokes number of the dust i-species.  

Following the same approach described in appendix A of D16, the dust drift velocity of the $i$-species $v^{i}_{\text{drift}}$ is given by
\begin{equation}
v^{i}_{\text{drift}}= \left( \dfrac{A}{1+\left(\text{St}^{i}\right)^2} + \dfrac{2B\text{St}^{i}}{1+\left(\text{St}^{i}\right)^2} \right) \eta v_{\text{k}}, 
\label{eq3-sec2-1}
\end{equation}
where the headwind $\eta v_{\text{k}}$ is given by
\begin{equation}
  \eta v_{\text{k}}= \dfrac{1}{2\rho_{\text{g}}\Omega_{\text{k}}} \dfrac{dP_{\text{g}}}{dR}.
  \label{eq4-sec2-1}
\end{equation}
The quantities $A$ and $B$ are given by\footnote{We note that there is a typo in the signs of Eqs.~A.4 and A.5 of D16. If the headwind $\eta v_k$ is defined as in our Eq.~\ref{eq4-sec2-1}, the correct signs are the ones shown for A and B in our Eqs.~\ref{eq5-sec2-1} \citep[see][]{Dipierro2018,Garate2020}.}
\begin{eqnarray}  \label{eq5-sec2-1}
  A &=& - \dfrac{2Y}{(1+X^2)+Y^2}, \nonumber \\ 
  \\ 
  B &=& \dfrac{1+X}{(1+X^2)+Y^2}, \nonumber
 \end{eqnarray}
where the dimensionless quantities $X$ and $Y$ are \citep{Dipierro2018, Garate2020} 
\begin{eqnarray}
  X &=& \sum \dfrac{\epsilon(m)}{1+\text{St}(m)^2}, \nonumber \\ 
  \\
  Y &=& \sum \dfrac{\epsilon(m)\text{St}(m)^2}{1+\text{St}(m)^2}. \nonumber
  \label{eq6-sec2-1}
\end{eqnarray}
As in D16, we use a power-law for the dust size distribution (or Stokes number distribution) given by
\begin{equation}
  n(\text{St}) \, m \, \text{d}\text{St} \propto \text{St}^{-1/2} \, \text{d}\text{St},
 \label{eq7-sec2-1}
\end{equation}
which implies a dust mass distribution $n(m) \propto m^{-11/6}$ where most of the dust mass remains in the biggest dust particles \citep{Birnstiel+2011}. Then, the mass weighted mean dust drift velocity results in
\begin{equation}
 \overline{v}_{\text{drift}}= \dfrac{ \sum_i \epsilon_i v_{\text{drift}}^i }{ \sum_i \epsilon_i }. 
 \label{eq8-sec2-1}
\end{equation}
In order to calculate $D^*$ and $\overline{v}_{\text{drift}}$ to solve Eq.~(\ref{eq1-sec2-1}), as in D16 we use 200 dust size bins between $r_{\text{d}}^0= 1~\mu\text{m}$ and $r_{\text{d}}^{\text{max}}$ for each radial bin. 

The initial dust surface density is given by 
\begin{equation}
    \Sigma_{\text{d}}(R) = \eta_{\text{ice}} z_{0} \Sigma_{\text{g}}(R) 
    \label{eq:eq8-sec2-1}
\end{equation}
where $\eta_{\text{ice}}$ takes into account the sublimation of water-ice given by
\begin{eqnarray}
   \eta_{\text{ice}}= 
   \begin{cases}
     1 & \text{ if $R \ge R_{\text{ice}}$},  \\
     \\
     {\dfrac{1}{\beta}} & \text{ if $R < R_{\text{ice}}$},
   \end{cases} 
   \label{eq:eq9-sec2-3}
\end{eqnarray}   
with $R_{\text{ice}} \sim 3$~au the often call ice-line, $\beta=2$ \citep{Lodders2003}, and $z_{0}= 0.01$ as the classical initial dust-to-gas ratio.

Finally, Eq.~(\ref{eq1-sec2-1}) is solved using an implicit Donor cell algorithm considering zero density as boundary conditions, and the time-step is controlled not allowing changes greater than 1\% in the dust surface density for each radial bin between consecutive models.    

\subsection{Planetesimals formation by streaming instability}
\label{sec:sec2-2}

The streaming instability \citep{Youdin&Goodman2005,Johansen2007} can be a possible mechanism for the spontaneous formation of km-sized planetesimals due to the clumping of radially drifting small particles. However, this process is not triggered along the entire protoplanetary disk, but in specific regions where, due to different mechanisms such as e.g, dead zones, photoevaporation and vortices; dust pile-ups occur \citep{Drazkowska+2016,Drazkowska+2017, Ercolano2017}. 

Following \citet{Drazkowska+2014} and D16, we implement a similar model of planetesimal formation by checking for every radial bin and at every time-step if the density of large pebbles (with $\text{St} > 10^{-2}$) is comparable to the gas density:
\begin{eqnarray}
   \sum_{\text{St}>10^{-2}}{\dfrac{\rho_{\text{d}}(\text{St})}{\rho_\text{g}}} > 1.
   \label{eq:eq1-sec2-4}
\end{eqnarray} 
At the locations (and times) where this condition is fulfilled, part of the dust surface density is transferred to a planetesimal surface density following
\begin{eqnarray}
   \text{d}\Sigma_{\text{p}} = \zeta\dfrac{\Sigma_{\text{d}}(\text{St}>10^{-2})}{T_{\text{K}}}\text{d}t,
   \label{eq:eq2-sec2-4}
\end{eqnarray}
where $\zeta = 10^{-3}$ is the planetesimal formation efficiency parameter \citep{Drazkowska+2017}, and $T_{\text{K}}$ the orbital period. A validation of the dust growth, dust evolution, and planetesimal formation in our model, comparing our results with the fiducial ones of D16, is presented in Appendix~\ref{apex-1}.  

The time evolution of the planetesimal surface density $\Sigma_{\text{p}}$ is calculated by solving the advection-diffusion Eq. \ref{eq1-sec2-1}, like with the pebbles. The time evolution of planetesimal eccentricities and inclinations (and the corresponding drift velocities, planetesimal-planet relative velocities and the planetesimal accretion rates) are calculated as in \citet{Guilera2014}, considering the stirring due to the gravitational perturbations by the planet and the damping due to gas drag. Because the formed planetesimals are assumed to be of 100~km radii, their Stokes numbers are much greater than unity. Thus, the planetesimal distribution is always in the Quadratic drag regime (instead on the Epstein regime), and the contribution of the diffusion term in the advection-diffusion equation is negligible.

\subsection {Planet growth and migration}
\label{sec:sec2-3}

Once the solid mass accumulated at the pressure bump in the form of planetesimals is enough to form a Moon-mass embryo, we put an embryo of this mass at the pressure bump location \citep{Liu2019}. Then, the embryo grows by the concurrent hybrid accretion of pebbles and planetesimals, and the surrounding gas.

In the case of pebbles (particles with $\text{St} \le 1$), the pebble accretion rate is given by \citet{Lambrechts+2014}:
\begin{eqnarray}
  \frac{d\text{M}_{\text{C}}}{dt}\bigg|_{\text{pebbles}} =
  \begin{cases}
     2 \Sigma_{\text{d}}(a_{\text{P}}) \text{R}_{\text{H}}^2 \Omega_{\text{P}} \cdot f_{\text{reduc}}, \text{ if}~ \, 0.1 \le $\text{St}$ < 1, \\
    \\
     2 \left(\frac{\text{St}}{0.1} \right)^{2/3} \Sigma_{\text{d}}(a_{\text{P}}) \text{R}_{\text{H}}^2 \Omega_{\text{P}} \cdot f_{\text{reduc}}, \text{ if}~$\text{St}$ < 0.1, 
  \end{cases}  
  \label{eq1-sec2-3}
\end{eqnarray}
where $\text{M}_{\text{C}}$ is the mass of the core, and $f_{\text{reduc}}= \text{min}(1, \text{R}_{\text{H}}/H_p)$ takes into account a reduction in the pebble accretion rate if the scale height of the pebbles, $\text{H}_p= \text{H}_{\text{g}}\sqrt{\alpha/(\alpha + \text{St})}$, becomes greater than the Hill radius of the planet. The planet can accrete pebbles until it reaches the usually called pebble isolation mass \citep{Lambrechts+2014}. When the planet becomes massive enough, the perturbations it generates onto the gaseous disk develop a partial gap, and the planet produces a pressure bump beyond its orbit. This pressure bump, different from the one generated by the viscosity transition at the ice-line, also acts as a dust trap. When this happens the accretion of pebbles by the planet is limited. Thus, when the planet reaches the pebble isolation mass given by \citep{Lambrechts+2014}
\begin{equation}
  \text{M}_{\text{iso}}^{\text{pebble}}= 20 \left( \dfrac{a_{\text{P}}}{5~\text{au}} \right)^{3/4} \text{M}_{\oplus},
  \label{eq2-sec2-3}
\end{equation}
we consider that the pebble accretion rate is halted ($a_{\text{P}}$ represents the planet location). We note, however, that pebble accretion has not been studied yet in a density maximum of a protoplanetary disk, therefore it is uncertain how big a solid core can grow by pebble accretion. It may also happen that a core can grow larger than the isolation mass determined in disks where the surface density profile is described by a radial power law profile \citep{Sandor2020}. Therefore in Section \ref{sec:iso} we also show a simulation in which the pebble isolation mass is lifted up.

We note that in order to compute Eq.~\ref{eq1-sec2-3} we use the the mass weighted mean Stokes number of the dust size distribution $\overline{\text{St}}$. 

In the case of planetesimals we use, as in our previous works, the planetesimal accretion rate for the oligarchic growth derived by \citet{Inaba2001}
\begin{equation}
  \dfrac{d\text{M}_{\text{C}}}{dt}\bigg|_{\text{plts}} = \Sigma_{\text{p}}(a_{\text{P}}) \text{R}_{\text{H}}^{2} P_{\text{coll}} \Omega_{\text{P}}, 
  \label{eq3-sec2-3}
\end{equation}
where $\Sigma_{\text{p}}(a_{\text{P}})$ is the surface density of planetesimals at the location of the planet, $\text{R}_{\text{H}}$ is the Hill radius of the planet and $P_{\text{coll}}$ is the collision probability. $P_{\text{coll}}$ depends on the planetesimal-planet relative velocities and in the enhanced capture radius of the planet due to the gaseous envelope \citep{Inaba&Ikoma-2003}. 

For the growth of the envelope, we compute the gas accretion solving the standard stellar evolution equations as in \citet{Guilera2014} but considering that the energy released by the solid accretion is strictly deposited in the core. In addition, in Appendix~\ref{apex-2} we compare the results from the fiducial simulations against the results computing the gas accretion adopting the prescriptions from \citet{Ikoma2000}. We note that if the gas accretion rate exceeds the amount that can be provided by the disk ($\dot{\text{M}}_{\text{disk}}= 3 \pi \nu \Sigma_{\text{g}}$ calculated at the planet's location), the gas accretion onto the planet switches to the disk supply. 

Regarding the migration of the planet, if the planet is not massive enough to open a gap in the gaseous disk, we consider that the torque exerted by the gaseous disk onto the planet is given by
\begin{equation}
  \Gamma= \Gamma_{\text{type I}} + \Gamma_{\text{thermal}},
  \label{eq4-sec2-3}
\end{equation}
where $\Gamma_{\text{type I}}$ represents the classical torque associated to the type I migration (Lindblad and corotation torques), and $\Gamma_{\text{thermal}}$ takes into account the heat released by the planet due to the solid accretion \citep{Masset2017}. For consistency, as in \citet{Guilera&Sandor2017}, we consider $\Gamma_{\text{type I}}$ for isothermal disks from \citet{Tanaka+2002} the fiducial simulations, 
\begin{eqnarray}
\dfrac{\Gamma_{\text{type I}}}{\Gamma_0}= (1.364 + 0.541 \alpha'), 
\label{eq4-1-sec2-3}
\end{eqnarray}
being $\alpha'= \text{d} \log \Sigma_{\text{g}} / \text{d} \log R$ at the planet location, and $\Gamma_0=~(q/h)^2 \Sigma_{\text{g}}^2 a_{\text{P}}^4 \Omega_k^2$. where q is the planet and the central star mass ratio, and the aspect ratio $h$, the gas surface density and the Keplerian frequency are calculated at the planet location. We note that the normalized torque $\Gamma_{\text{type I}}/\Gamma_0$ only depends on the local gradient of the gas surface density.

For completeness, we also perform simulations adopting type I migration prescriptions derived for non-isothermal disks. In particular, we adopt also the migration recipes from \cite{paardekooper.etal2011} and \cite{jm2017}. For both recipes, the normalized torque $\Gamma_{\text{type I}}/\Gamma_0$ can be expressed by \citep[see][for details]{Guilera+2019}
\begin{eqnarray}
\dfrac{\Gamma_{\text{type I}}}{\Gamma_0}= \Gamma_L + \Gamma_C,
\label{eq4-2-sec2-3}
\end{eqnarray}
where $\Gamma_L$ is the Lindblad torque and $\Gamma_C$ is the corotation torque. In both cases, the Lindblad torque can be expressed as $\Gamma_L= c_1 + c_2 \beta' + c_3 \alpha'$, being $\beta'= \text{d} \log \text{T} / \text{d} \log R$ the local disk temperature gradient at the planet location. The constants $c_1, c_2$ and $c_3$ slightly differ between both recipes. Regarding the corotation torque component, its expression is more complex for both prescriptions. It depends mainly on the mass of the planet, the width of the corotation region, and the properties of the disk, such as the viscosity, opacity, thermal diffusivity, etc. \citep[see appendix of][for details]{Guilera+2019}. As it was shown in \citet{Guilera+2019}, the contribution of the corotation torque is where both migration recipes present more differences.

To compute $\Gamma_{\text{thermal}}$ we use the same methodology as in \citet{Guilera+2019}, where $\Gamma_{\text{thermal}}$ is given by \citep{Masset2017}
\begin{eqnarray}
\dfrac{\Gamma_{\text{thermal}}}{\Gamma_0}= 1.61 \dfrac{\gamma_{\text{g}}-1}{\gamma_{\text{g}}} \eta' \left( \dfrac{\text{H}_{\text{g}}}{\lambda_{\text{c}}} \right) \left( \dfrac{\text{L}}{\text{L}_{\text{c}}} -1 \right),
\label{eq4-3-sec2-3}
\end{eqnarray}
where $\eta'= \alpha'/3 + \beta'/6 + 1/2$, and $\lambda_{\text{c}}= \sqrt{\chi/q\Omega_k\gamma_\text{g}}$, being $\chi$ the gas thermal diffusivity at the location of the planet. $\text{L}$ is the luminosity of the planet due to solid accretion and $\text{L}_{\text{c}}$ represents a critical luminosity \citep[see][for details]{Guilera+2019}. We remark that when the mass of the planet becomes greater than the critical thermal mass, defined by \citep{Masset2017}
\begin{equation}
\text{M}_{\text{crit}}^{\text{thermal}}= \dfrac{\chi c_s}{G},
\label{eq5-sec2-3}
\end{equation}
with $G$ the  gravitational constant, the thermal torque is halted.  

As we are interested in the early formation of a proto-giant planet, for simplicity we stop our simulations when the planet becomes massive enough to open a gap in the disk following the criterion given by \citet{Crida+2006}. Thus, type II migration is not considered in this work. 

\section{Results}
\label{sec:sec3}

In the following sections we first compute the gas and dust disk evolution without considering planet formation and describe the main characteristics of the whole process, including dust growth and planetesimal formation by streaming instability. Then we include the planet formation at the pressure maximum by the hybrid accretion of pebbles and planetesimals.

\subsection {Disk evolution without planets}
\label{sec3-1}
 
At first we just compute the dust evolution, dust growth and the formation of planetesimals in the regions where the conditions for the streaming instability are fulfilled, but without placing any embryo. 
 
\begin{figure}
  \centering
  \includegraphics[width= 0.475\textwidth]{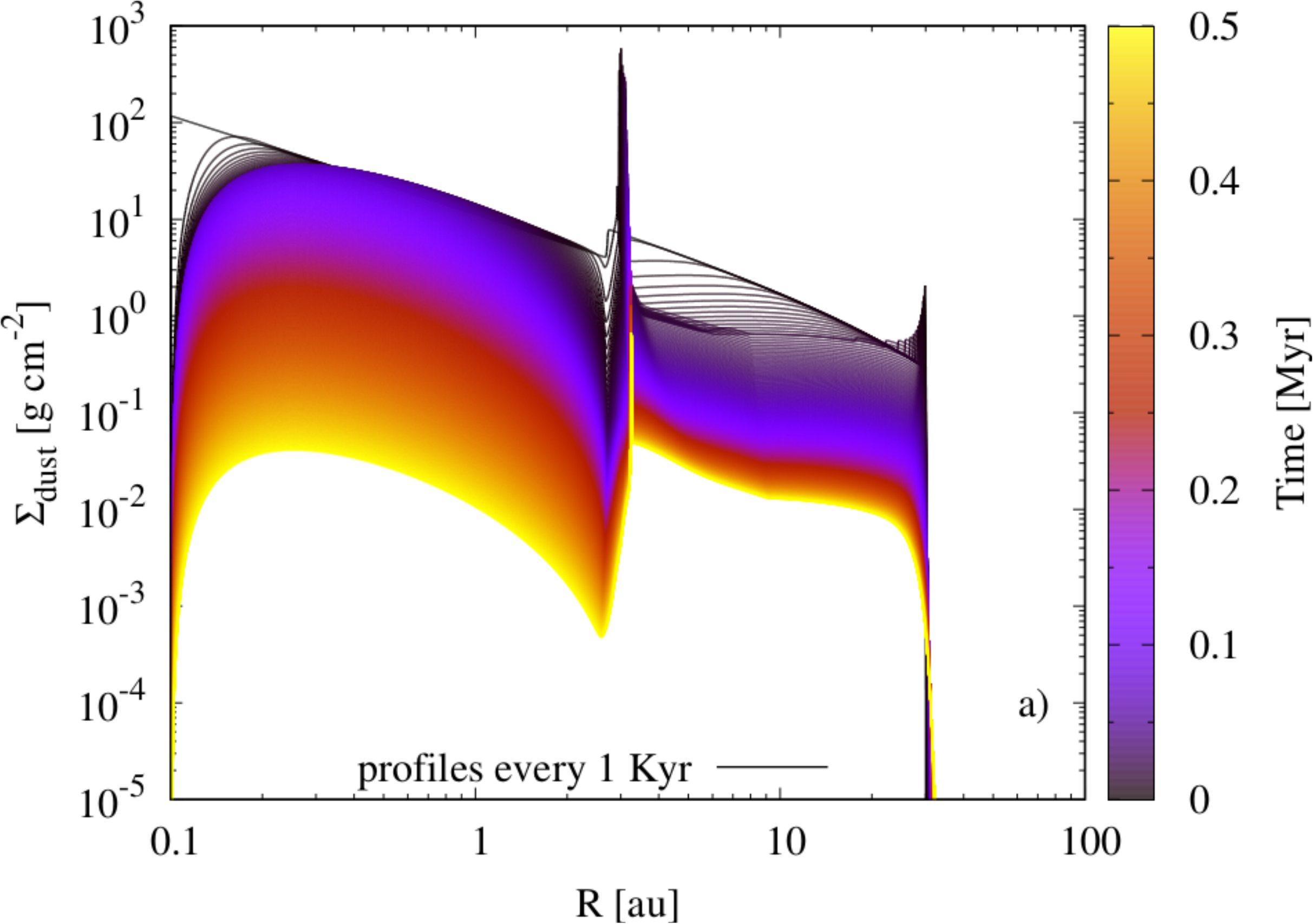} \\
  \includegraphics[width= 0.475\textwidth]{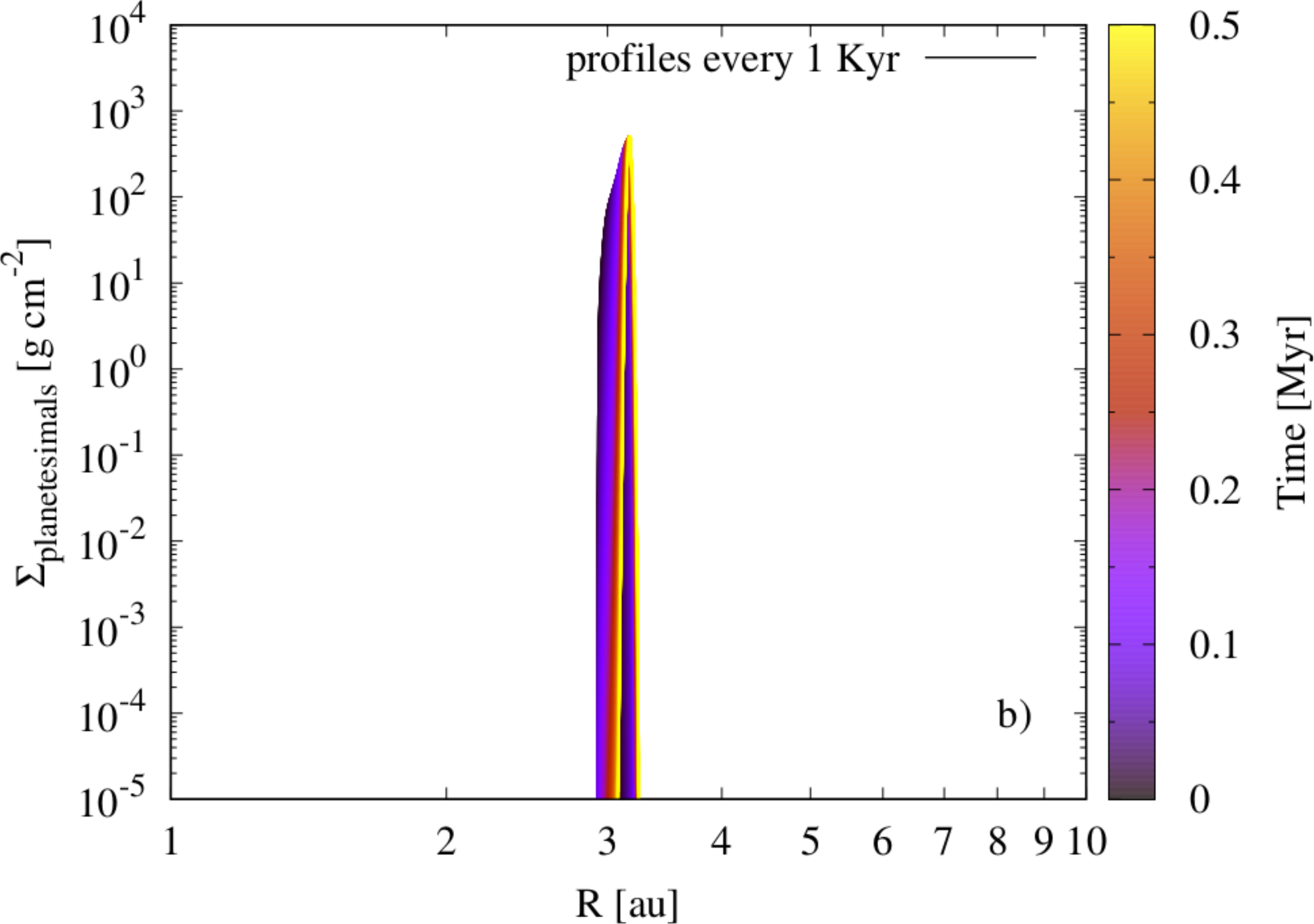}
  \caption{Time evolution of the radial profiles of the surface density of dust (panel a) and planetesimals (panel b). Planetesimals are only formed near the pressure maxima location. This simulation corresponds to the case of the disk evolution without planet formation.}
  \label{fig1-sec3-1}
\end{figure}

In Fig.~\ref{fig1-sec3-1} we show the time evolution of the radial profiles of the surface density of dust and planetesimals using the same disk parameters that the ones described in Sec.\ref{sec:sec2-0}. We note that the dust is very efficiently accumulated at the pressure maximum located at $\sim 3$~au (see Fig.~\ref{fig1-sec2-0}). This is a natural consequence of the null value of the pressure gradient and dust drift velocity at that location. The surface density of dust increases about two orders of magnitude at this position, allowing the formation of planetesimals by the streaming instability. Since the pressure maximum moves slightly outwards in time, as it was noticed in \citet{Guilera&Sandor2017}, this region is a narrow ring of  $\sim 0.3$~au and not a fixed unique value of semi-major axis. In contrast to \citet{Guilera&Sandor2017}, the dust diffusion which is now considered in Eq. \ref{eq1-sec2-1}, does not allow a significant dust accumulation at the inner edge of the disk (at $\sim 0.1$~au) and a significant amount of dust is lost inside the ice-line \citep[see Fig.~15 of][]{Guilera&Sandor2017}. In contrast to D16, we find that the condition for the streaming instability is not fulfilled in the inner part of the disk. This is due to the fact that the fragmentation threshold velocity considered in D16, which is constant along the disk, is much higher (between 8~m/s and 15~m/s) than the one used here for pure silicate dust. A similar result was found in \citet{Drazkowska+2017}.
 
\begin{figure}
  \centering
  \includegraphics[width=0.475\textwidth]{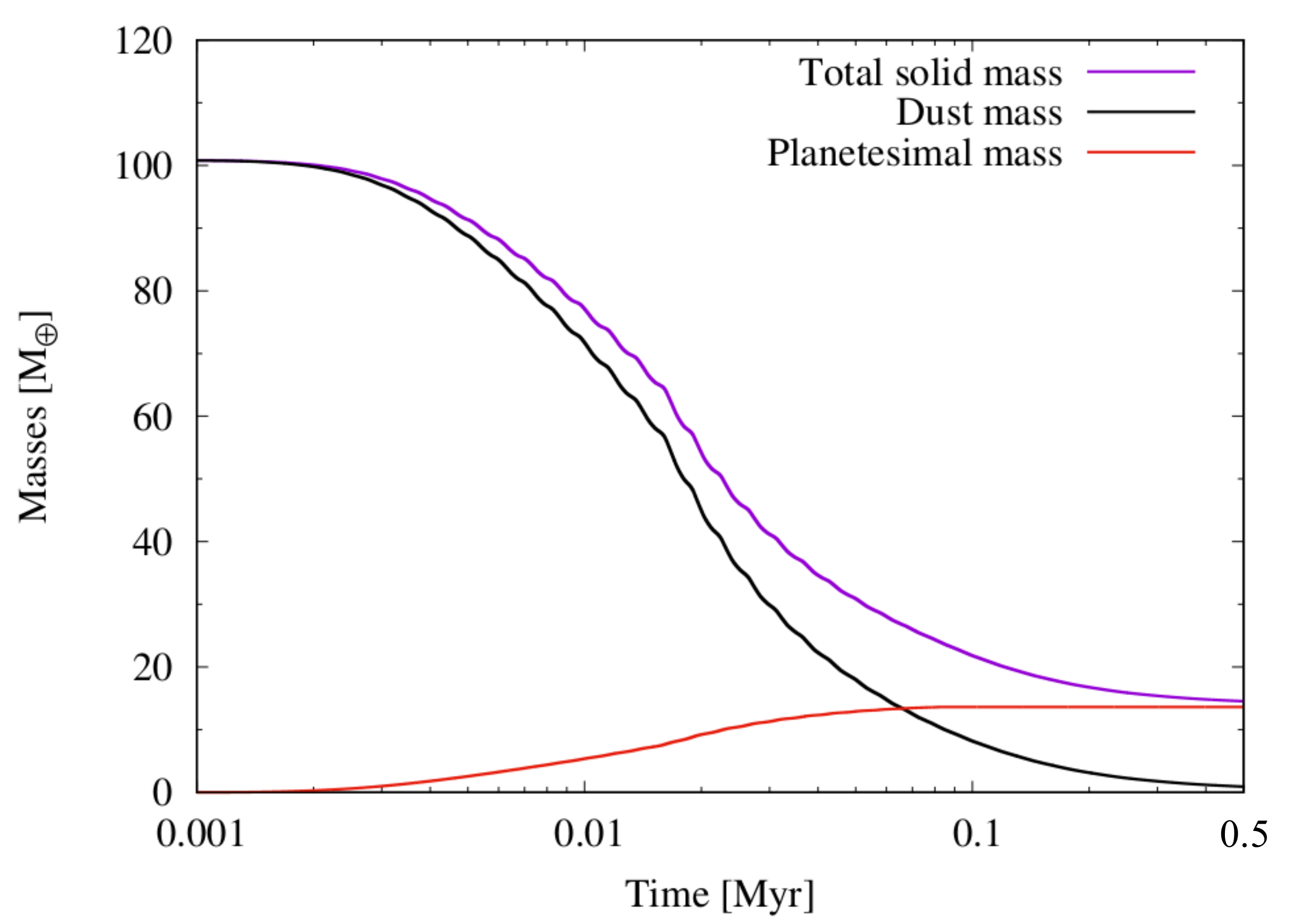} 
  \caption{Time evolution of the total mass of solids, dust and planetesimals. In just $5\times10^5$~yr there is almost no more dust in the disk. The simulation corresponds to the case of the disk evolution without planet formation.}
  \label{fig2-sec3-1}
\end{figure}

Figure~\ref{fig2-sec3-1} shows the time evolution of the total mass of solids in the disk. Initially, all the solid mass of the disk remains in dust particles of 1 $\mu$m. The disk has initially $\sim100 \, \text{M}_{\oplus}$, and as time evolves the dust mass decreases due to dust diffusion and accretion onto the star. The efficient accumulation of dust at the pressure maximum allows the formation of planetesimals by streaming instability at this location in a few thousand years. In about $10^5$~yr, $\sim17.5\text{M}_{\oplus}$ of dust are transformed into planetesimals. After that no further planetesimals form and the planetesimal mass remains basically constant. This is due to the significant decrease in the surface density, flux and mass of pebbles, and the mass of dust which continues decreasing by dust diffusion and accretion onto the star. 

\begin{figure}
  \centering
  \includegraphics[width= 0.475\textwidth]{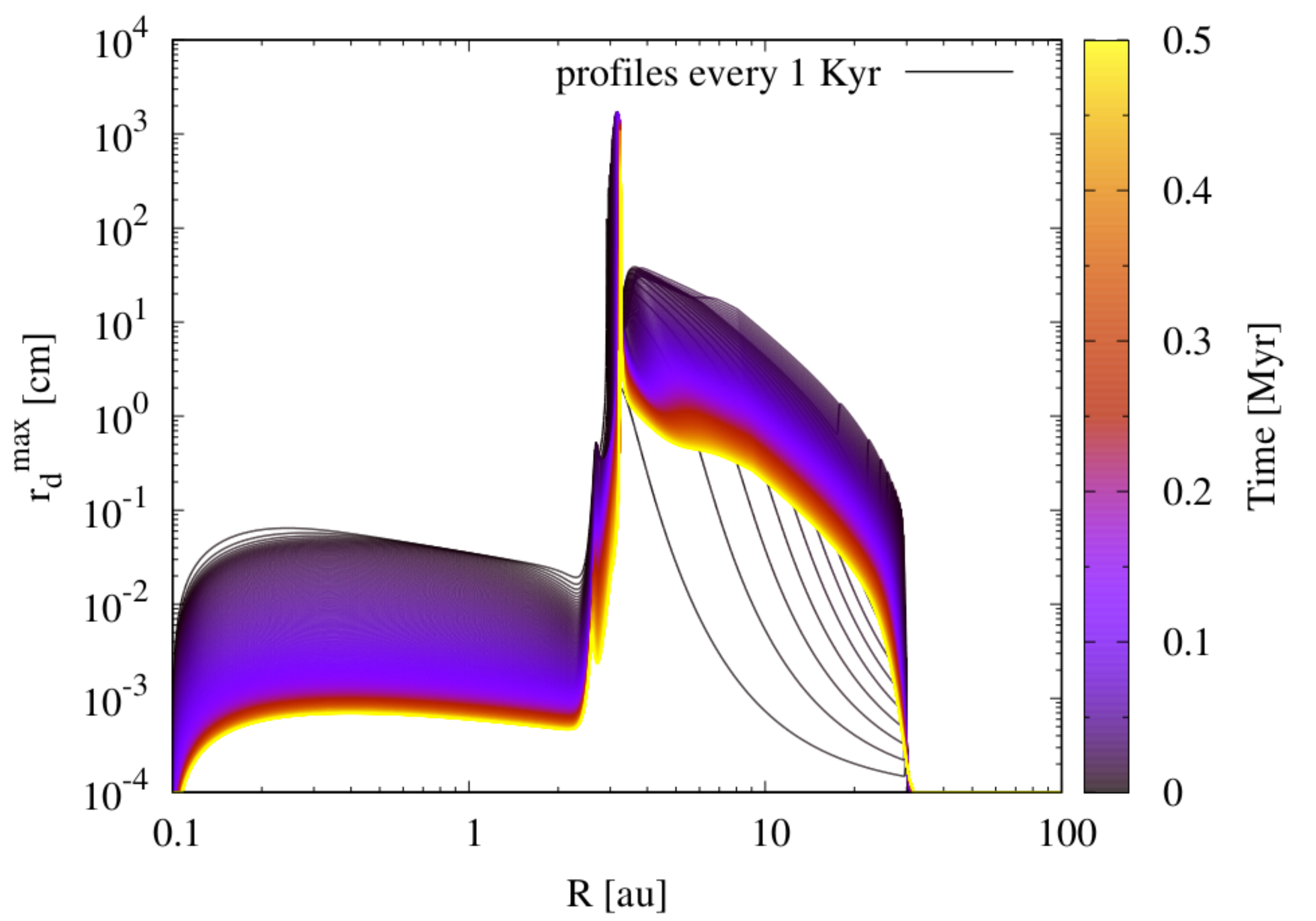} 
  \caption{Time evolution of the radial profiles of the maximum sizes that the dust can reach. The maximum sizes for the dust population are reached at the pressure maximum location. The simulation corresponds to the case of the disk evolution without planet formation.}
  \label{fig3-sec3-1}
\end{figure}

In Fig.~\ref{fig3-sec3-1}, we show the time evolution of the maximum dust size that the dust population can reach along the disk. Due to the fact that beyond the ice-line, the threshold velocity for ice-rich dust is higher than in the inner part of the disk, dust can grow to larger sizes. In particular, near the pressure maximum, where the gas and dust surface densities are higher, the particles reach sizes of $\sim 10$~m ($r^{\text{max}}_{\text{drift}}$, $r^{\text{max}}_{\text{drift}}$ and $r^{\text{max}}_{\text{ddf}}$ from Eq. \ref{eq1-sec2-1-0} are proportional to $\Sigma_\text{d}$ and $\Sigma_\text{g}$). In the inner part of the disk, the dust can only grow to sizes up to $\sim$ 0.1 cm. 

\begin{figure}
  \centering
  \includegraphics[width= 0.475\textwidth]{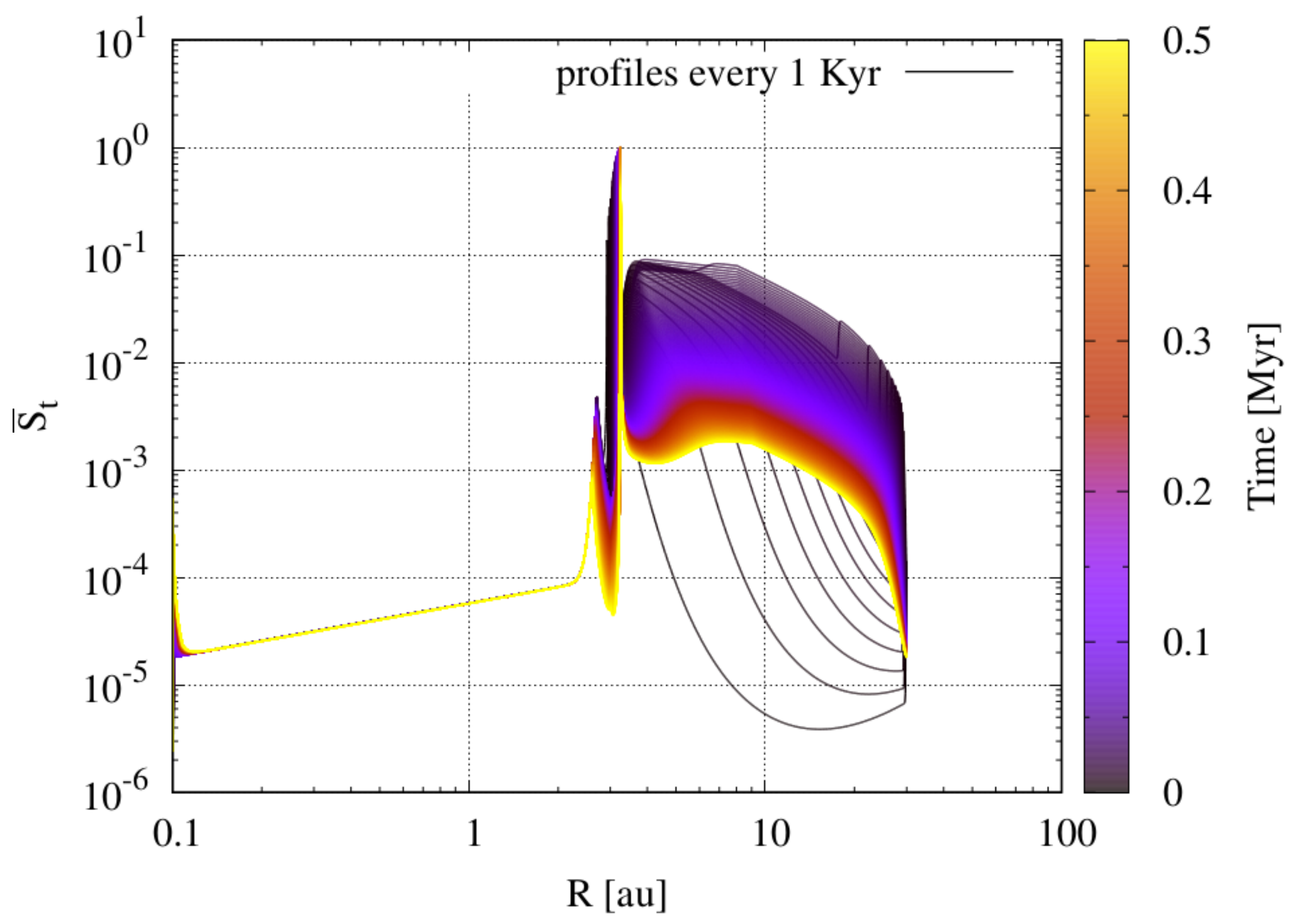} 
  \caption{Time evolution of the radial profiles of the mass weighted mean Stokes number of the dust size distribution. As for the dust sizes, the maximum values are reached at the pressure maximum location. The simulation corresponds to the case of the disk evolution without planet formation.}
  \label{fig4-sec3-1}
\end{figure}

Fig.~\ref{fig4-sec3-1} shows the time evolution of the corresponding mass weighted mean Stokes number of the dust population along the disk. The mean Stokes number at the pressure maximum reaches values near unity, higher than the $\text{St} > 0.01$ needed to fulfill the streaming instability condition, allowing the formation of planetesimals.

 We note that as we consider that the disk temperature does not evolve in time, the location of the viscosity transition does not change in time. In general, in non-isothermal disks the temperature decays in time and the ice-line moves inward. So, in more realistic disks the location of the viscosity transition could also moves inward in time. However, despite our simplification, we showed that the efficient accumulation of dust at the pressure maximum generated at the viscosity transition allows the formation of planetesimals in a short period of time. Thus, our model can be a good approximation if the time-scale of the drift of the ice-line is not much shorter than the one corresponding to the formation of planetesimals. 
 
We also note that, for simplicity, the disk is initially cut at 30 au. The initial outer disk edge defines the initial disk mass and regulates the time evolution of the pebble flux, thus, it could play an important role. However, this is not the case in this work since the pressure bump at the ice-line efficiently collects the inward drifting pebbles and the pebble isolation mass is quickly reached by the planet (see next section). 
  
 \subsection{Giant planet formation at the pressure maximum location}
\label{sec3-2}
 
As we mentioned before, once the mass in planetesimals reaches the mass of the Moon, we transform this mass into an embryo and insert it at the location of the pressure maximum, where planetesimals form. After that, the flux of pebbles of the outer region of the disk continues being accumulated at the pressure bump and pebbles continue forming planetesimals (when the SI criterion is fulfilled). Thus, the embryo grows by the concurrent accretion of pebbles, planetesimals and the surrounding gas.
  
\begin{figure}
  \centering
  \includegraphics[width= 0.475\textwidth]{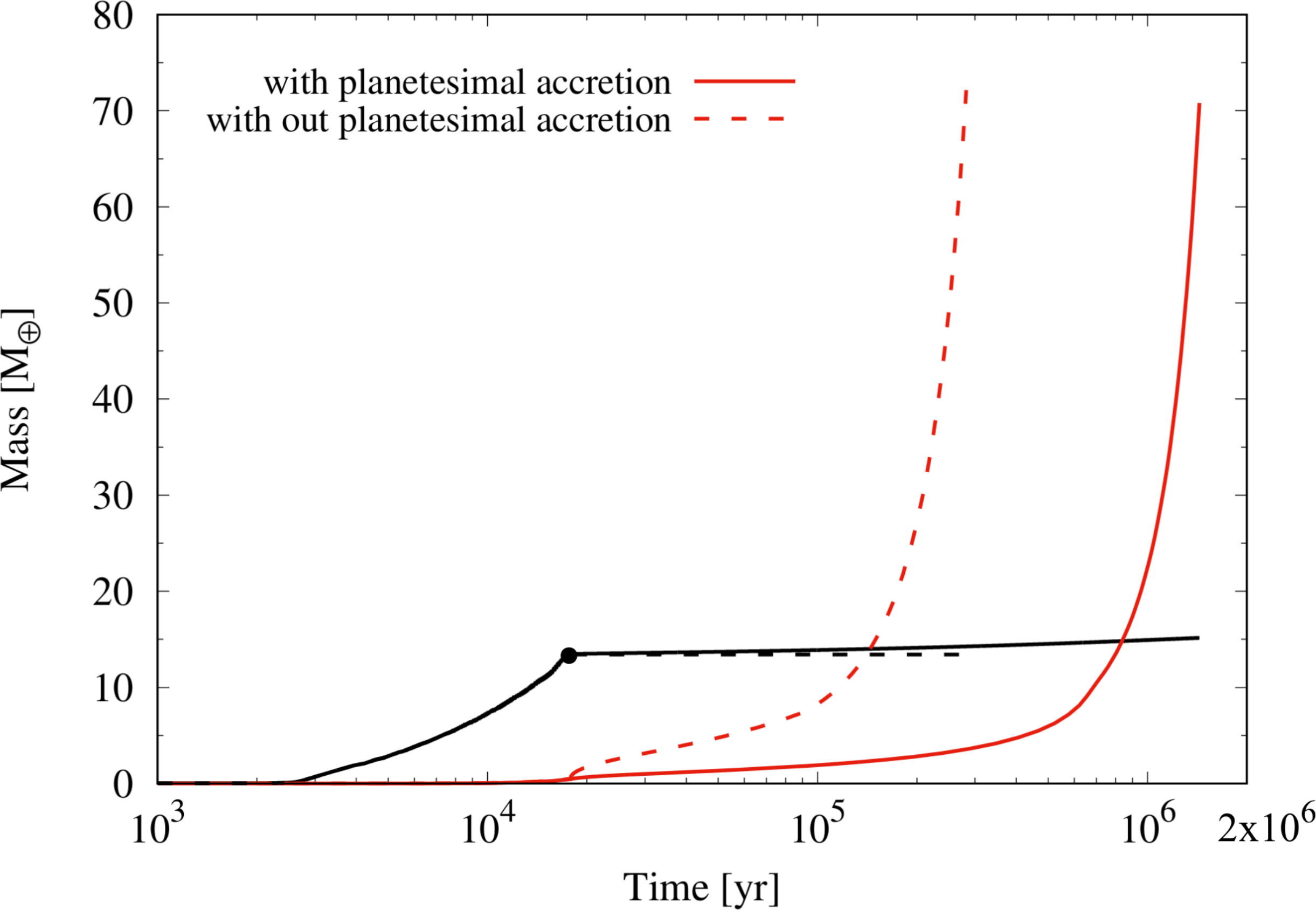} 
  \caption{Time evolution of the mass of the core (the black lines) and the envelope mass of the planet (the red lines). The solid lines represent the model where the hybrid accretion of pebbles and planetesimals is considered. The dashed lines correspond to the case where the accretion of planetesimals after the planet reaches the pebble isolation mass (black filled dot) is neglected.} 
  \label{fig1-sec3-2}
\end{figure}

In Fig.~\ref{fig1-sec3-2} we show the time evolution of the mass of the core (black lines) and the mass of the envelope of the planet (red lines) for two different simulations. In these simulations the embryo initially grows mainly due to accretion of pebbles, basically because the surface density and flux of pebbles dominate. Due to the significant amount of pebbles accumulated at the density/pressure maximum, the planet reaches the pebbles isolation mass (of $\sim 13.5~\text{M}_{\oplus}$) at only $\sim 2\times10^4$~yr. 

 In order to analyze the gas accretion, we first compute, for the sake of simplicity, a simulation where we neglect planetesimal accretion once the pebble isolation mass is reached. The growth of the envelope for this simulation is represented by the red dashed line (Fig.~\ref{fig1-sec3-2}). To compute the gas accretion we use our full model, i.e we solve the envelope structure and transport equations via a Henyey method \citep{Benvenuto2005}. We show that if solid accretion suddenly stops, gas is accreted very efficiently and the formation of a gas giant occurs in a time-scale of $\sim 10^5$ years. In the case presented here, the cross-over mass (when the mass of the envelope equals the mass of the core, and runaway gas accretion is triggered) is reached at $\sim 1.5 \times 10^5$~yr, and the simulation stops when the planet opens a gap in the disk at $\sim 3 \times 10^5$~yr. This happens because once solid accretion is halted, there is not enough heat to prevent the fast compression of the envelope, which boosts the accretion of gas.
 
\begin{figure}
  \centering
  \includegraphics[width= 0.475\textwidth]{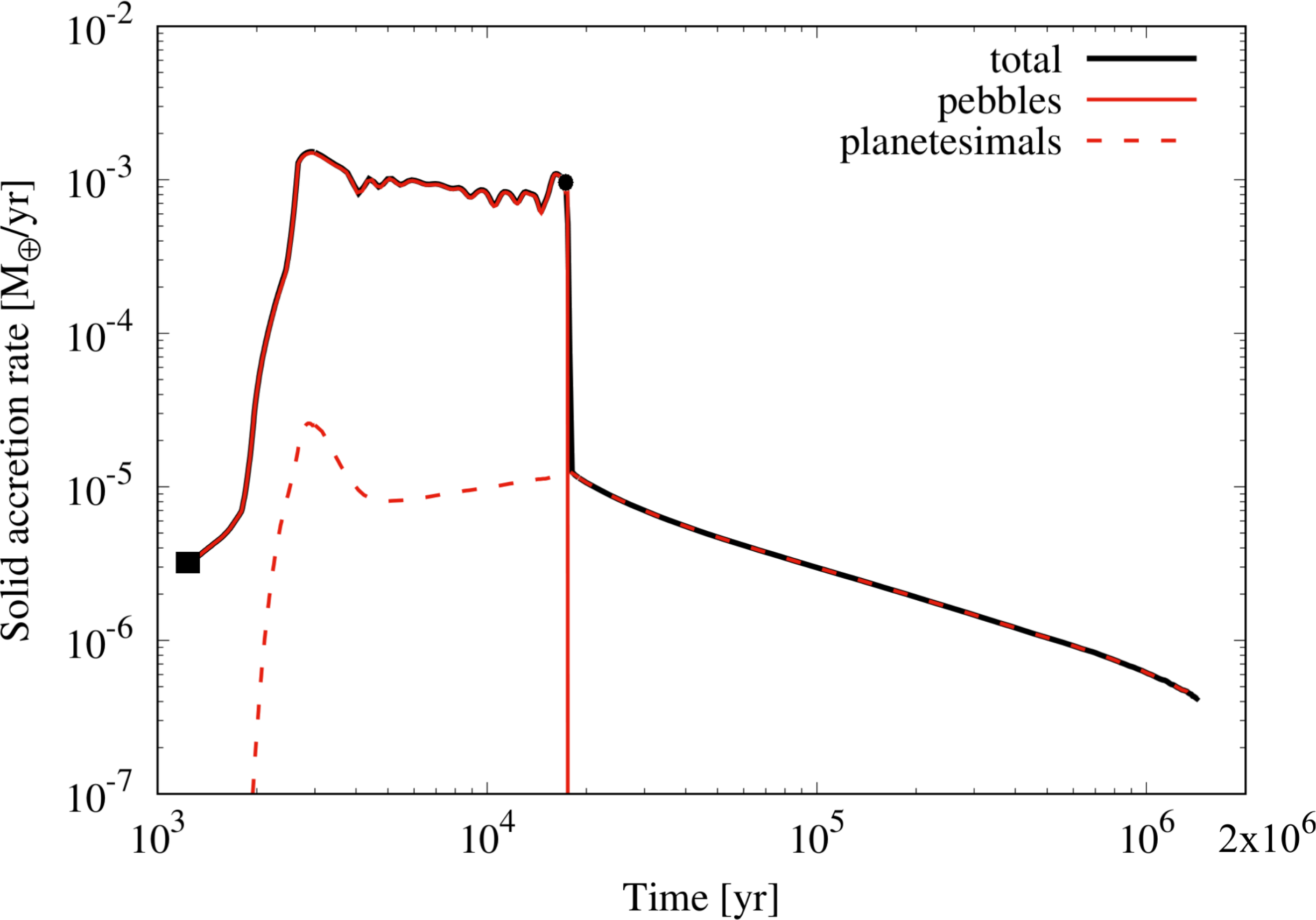} 
  \caption{Time evolution of the solid accretion rate for the simulation where the hybrid accretion of pebbles and planetesimals is considered. The black filled square corresponds to the time at which the embryo is introduced in the simulation. Initially, the total solid accretion rate is dominated by the accretion of pebbles (the solid red line). After the planet reaches the pebble isolation mass (black filled dot), the accretion rate decays significantly and is only due to the accretion of planetesimals (the dashed red line).}
  \label{fig2-sec3-2}
\end{figure}

The second simulation corresponds to the more realistic case where the accretion of planetesimals continues after the planet reaches the pebble isolation mass (red-solid line of Fig.~\ref{fig1-sec3-2}). We can see, in Fig.~\ref{fig2-sec3-2}, the time evolution of the solid accretion rate. The black square represents the time at which the embryo is inserted in the simulation. We can see that due to the significant accumulation of pebbles at the density/pressure maximum the solid accretion rate remains very high until the planet reaches the pebble isolation mass. After that, the solid accretion rate drops about two orders of magnitude and then it is due entirely to the accretion of planetesimals. We remark that in this simulation, after the planet reaches the pebble isolation mass, the formation of planetesimals is stopped. This is because at this point, the planet generates a partial gap in the gaseous disk and the flux of pebbles is halted. Thus, the planetesimal accretion rate decreases in time from $\sim 10^{-5}~\text{M}_{\oplus}/\text{yr}$ to $\sim 3\times10^{-7}~\text{M}_{\oplus}/\text{yr}$ by two factors: the accretion of planetesimals that decreases the planetesimal surface density and the increase in the relative velocities between the planet and the planetesimals due to the gravitational perturbations as the planet grows. 
 
Regarding the gas accretion in this last scenario, we clearly see that the accretion of planetesimals delays the onset of the gaseous runaway. The energy released by the accretion of planetesimals avoids the quickly compression of the envelope layers slowing down the accretion of gas. In this case, the planet opens a gap in the disk (and the simulation stops) when it reaches a total mass of $\sim 85~\text{M}_{\oplus}$ at about 1.5~Myr. This implies a delay of about 1.2~Myr compared to the case where the accretion of planetesimals is neglected.
  
\subsection{Effect of type I migration} 

\begin{figure}
  \centering
  \includegraphics[width= 0.475\textwidth]{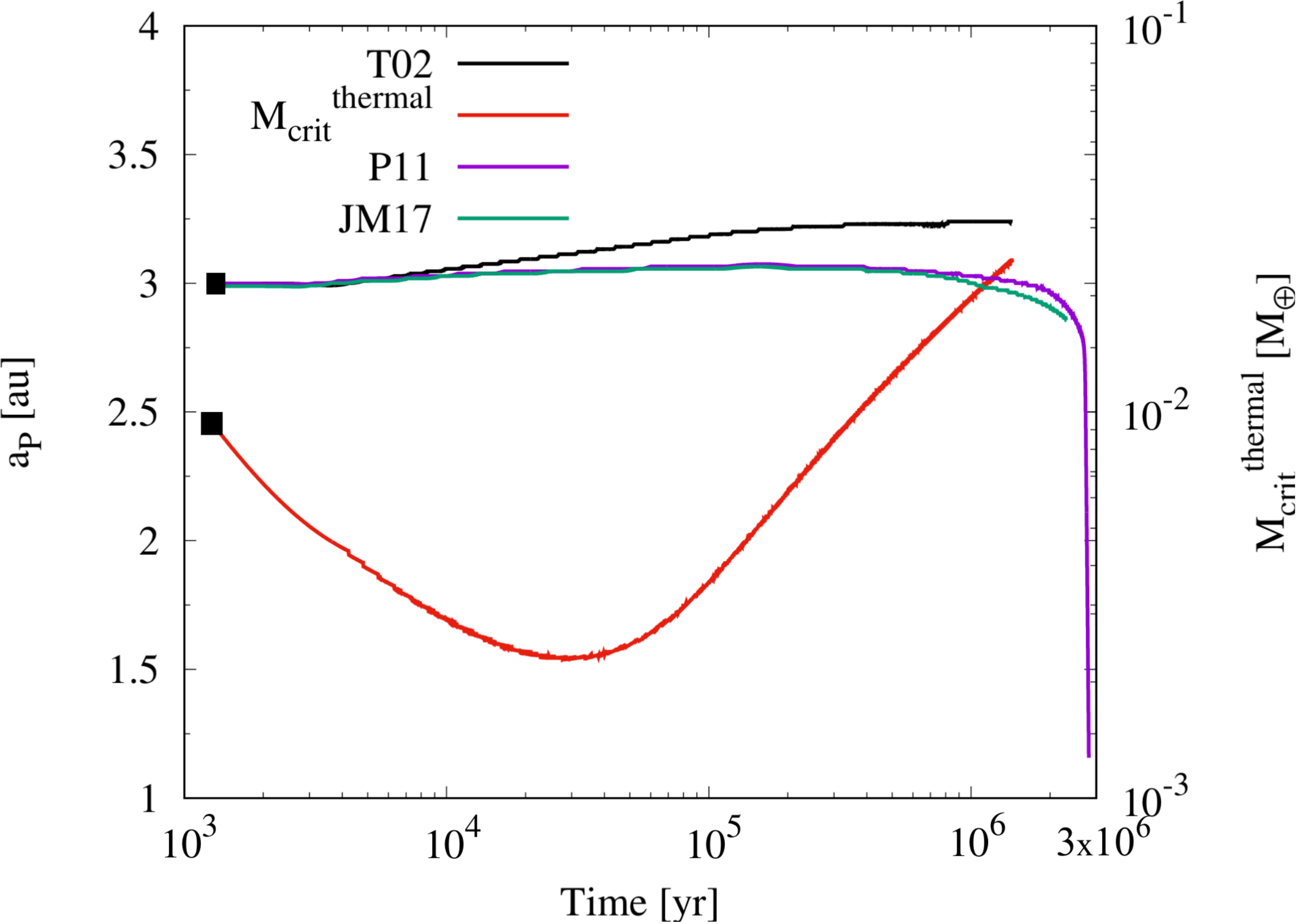} 
  \caption{Time evolution of the semi-major axis of the planet for the simulations corresponding to the case where the hybrid accretion of pebbles and planetesimals is considered. The black line corresponds to the fiducial simulation where type I migration recipes from \citet{Tanaka+2002} are adopted. The purple and green lines correspond to the cases where type I migration recipes from \citet{paardekooper.etal2011} and \citet{jm2017} are considered, respectively. For the cases T02 and JM17, the planet remains in a migration trap until it opens a gap in the gas disk. The red line represents the values of the critical thermal mass associated to the thermal torque (for the T02 case). As this mass is always much lower than the mass of the planet, the thermal torque does not play any effect in the torque over the planet.}
  \label{fig3-sec3-2}
\end{figure}

For the fiducial simulations described in the previous section, type I migration from \citet{Tanaka+2002} plus the thermal torque from \citet{Masset2017} was included. Figure \ref{fig3-sec3-2} shows the evolution of the semi-major axis of the planet and the critical thermal mass (see definition in Sec.~\ref{sec:sec2-3}) for the case where the hybrid accretion of pebbles and planetesimals is considered (black line). First, we can see that the critical thermal mass remains always lower than the mass of the planet. In fact, it only becomes greater than $0.01~\text{M}_{\oplus}$ (the initial mass of the embryo) at about $4\times10^5$~yr, after the planet reached the pebble isolation mass. Thus, the thermal torque is not contributing to the total torque onto the planet. We note that this situation could change if a more complex treatment of the thermodynamics of the disk is considered. However, the computation of the vertical structure in a non-isothermal disk with a viscosity transition at the ice-line presents some challenges. The main problem lies on the computation of the disk's vertical structure, which yields the mid-plane temperature, mean viscosity, etc. Those calculations require the previous knowledge of $\alpha$, and hence, of the exact location of the transition in $\alpha$, which also depends on the mid-plane temperature. Hence, some iterative procedure should be envisioned to tackle such problem, which is beyond the scope of this work. Moreover, the initial location of the ice-line depends also on the particular initial mass of the disk and the initial gas surface density. Thus, in this work we keep the simplified approach of an isothermal vertically disk with a temperature profile that does not evolve in time.

As in \citet{Guilera&Sandor2017}, we find that the density/pressure maximum acts as a planet migration trap for the fiducial simulation where the hybrid accretion of pebbles and planetesimals is considered. As we showed in \citet{Guilera&Sandor2017}, the density/pressure maximum generates a zero torque location near such maximum (at $\sim 3$~au). This zero torque location remains during the formation of the planet until it switches to type II migration since during type I migration the normalized torque for isothermal disks only depends on the local gradient of the gas surface density (see Eq.~\ref{eq4-1-sec2-3}). When this occurs, the planet migrates inward. Thus, the pressure maximum generated by the viscosity transition does not prevent the existence of gas giants in the inner region of the disk.
 
\begin{figure}
  \centering
  \includegraphics[width= 0.475\textwidth]{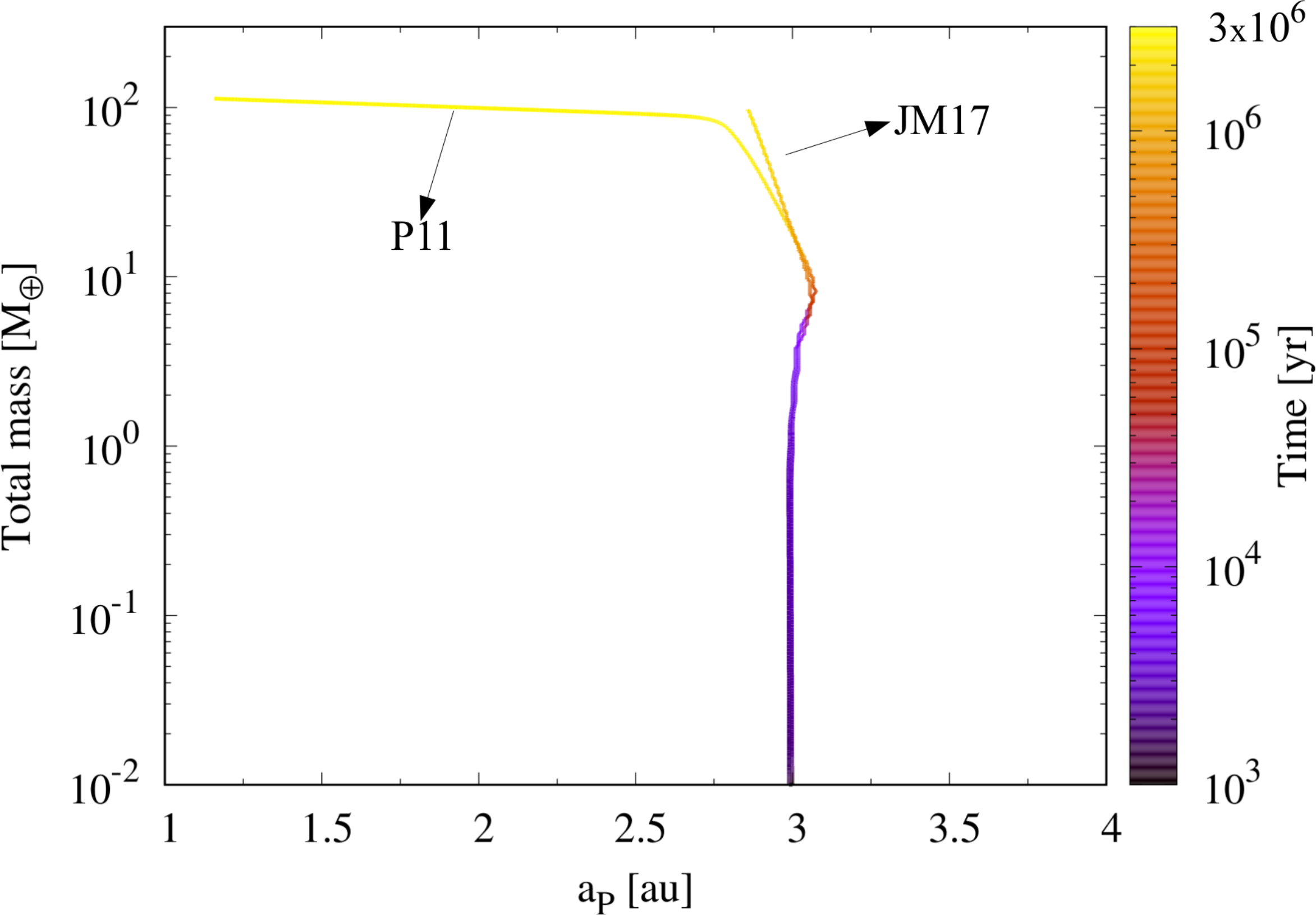} 
  \caption{Planet formation tracks for the simulations using the type I migration recipes from \citet{paardekooper.etal2011}(P11) and \citet{jm2017} (JM17).}
  \label{fig4-sec3-2}
\end{figure}

 In addition, we also perform two new simulations changing the type I migration recipes from \citet{Tanaka+2002} to \citet{paardekooper.etal2011} and \citet{jm2017}. With this, we want to analyze if the planet migration trap could be a robust result even using more sophisticated type I migration recipes, which were derived for non-isothermal disks. We remark that despite our simplifications in the computation of the thermodynamics of the disk, we compute the time evolution of the radial profiles of all the quantities needed to calculated the migration recipes mentioned above, i.e, we compute the time evolution of the radial profiles of the density, pressure, opacity, etc., while the temperature radial profile remains fixed in time. To be explicit, the simplification lies in that the evolution of the temperature is not linked self-consistently with the evolution of the gas surface density. We can see in Fig.~\ref{fig3-sec3-2} that the density/pressure maximum also acts as a planet migration trap using the type I migration recipes from \citet{jm2017}. However, using the type I migration prescriptions from \citet{paardekooper.etal2011}, the migration trap is broken close to the mass needed for the planet to open a gap in the disk. This happens because now the normalized torque is not only a function of the local gradient of the gas surface density (as in the isothermal case) but also of the local gradient of the temperature, the viscosity and the mass of the planet (these last two dependencies through the corotation torque). Thus, the combination of such quantities break the migration trap at some moment. \citet{Guilera+2019} showed that the main differences between the migration recipes from \citet{jm2017} and \citet{paardekooper.etal2011} lie on the computation on the corotation torque. Thus, in this case when the planet reaches a mass of a several tens of Earth masses, the migration trap is broken. At this moment, the planet quickly migrates inwards (due to the fact that it has a mass of $\sim 80~\text{M}_{\oplus}$) until it opens a gap at $\sim 1.15$~au. We remark that the time evolution of the mass of the core and the mass of the envelope for the simulations using the type I migration recipes from \citet{paardekooper.etal2011} and \citet{jm2017} are very similar to those of the fiducial simulation. In Fig.~\ref{fig4-sec3-2}, we emphasize this by plotting the planet formation tracks as a function of time. Simulations also stopped when the planets open a gap in the disk. This happens when the total mass of the planet is about 110~$\text{M}_{\oplus}$ for the case where we used the migration recipes from  \citet{paardekooper.etal2011} and about 95~$\text{M}_{\oplus}$ for the case where we adopt the migration recipes from \citet{jm2017}. The differences in the total mass at which the planet opens the gap are due to the fact that this mass depends on the viscosity of the disk. Inside the ice-line, the viscosity is larger (because of the larger $\alpha$-parameter), thus the total mass of the planet has to be higher to open a gap. For the second case, the planet opens the gap in the $\alpha$ transition region between $10^{-5}$ (outside the ice-line) and $10^{-3}$ (inside the ice-line). We note that if the thermodynamic of the disk is computed self-consistently, the migration trap could break at a different planet mass depending mainly on the viscosity and thermal diffusivity of the disk \citep[see][]{Morbidelli2020}.

\subsection{About the pebble isolation mass}
\label{sec:iso}
In the previous sections we adopted the pebble isolation mass estimated by \citet{Lambrechts+2014}. The pebble isolation mass is a very important quantity because it determines how massive a core can growth by pure pebble accretion. The maximum value that a core can reach has also an important role in the gas accretion rates. The more massive the core becomes, the higher are the gas accretion rates when the pebble accretion is halted \citep{Ikoma2000}. The pebble isolation mass strongly depends on the disk properties, mainly on the disk aspect ratio and on the disk viscosity, both at the location of the planet \citep[e.g][]{Bitsch2018}. Recently, \citet{Ataiee2018} studied the dependence of the pebble isolation mass with the disk viscosity and studied as well how the dust diffusion modifies the pebble isolation mass obtained from pure gas hydrodynamical simulations. For this last case, these authors find that pebble isolation mass is given by 
\begin{eqnarray}
  \text{M}_{\text{iso, gas}}^{\text{pebbles}}= \left( \dfrac{\text{M}_{\star}}{\text{M}_{\oplus}} \right) h_{\text{P}}^3 \sqrt{82.33\alpha_{\text{P}} + 0.03} \text{M}_{\oplus}, 
  \label{eq1-sec3-3}
\end{eqnarray}
where $\text{M}_{\star}$ is the mass of the central star (1~$\text{M}_{\odot}$ in this work), and $h_{\text{P}}$ and $\alpha_{\text{P}}$ are the disk aspect ratio and the $\alpha$-viscosity parameter at the planet location, respectively. They also studied how dust diffusion modifies the pebble isolation described above through hydrodynamic simulations of gas plus dust. For particles with Stokes numbers greater than 0.05 and a fixed aspect ratio, they find that the differences between both isolation masses depends on a factor greater than unity, which is also $\alpha$-dependent. For the lowest $\alpha$ value studied by \citet{Ataiee2018} ($\alpha= 5\times10^{-4}$), they find that this factor is 1.3. Thus, as a conservative approach, we take this value, meaning that the pebble isolation mass with dust diffusion is given by 
\begin{eqnarray}
  \text{M}_{\text{iso, dust}}^{\text{pebbles}}= 1.3~ \text{M}_{\text{iso, gas}}^{\text{pebbles}}.
  \label{eq2-sec3-3}
\end{eqnarray}

\begin{figure}
  \centering
  \includegraphics[width= 0.475\textwidth]{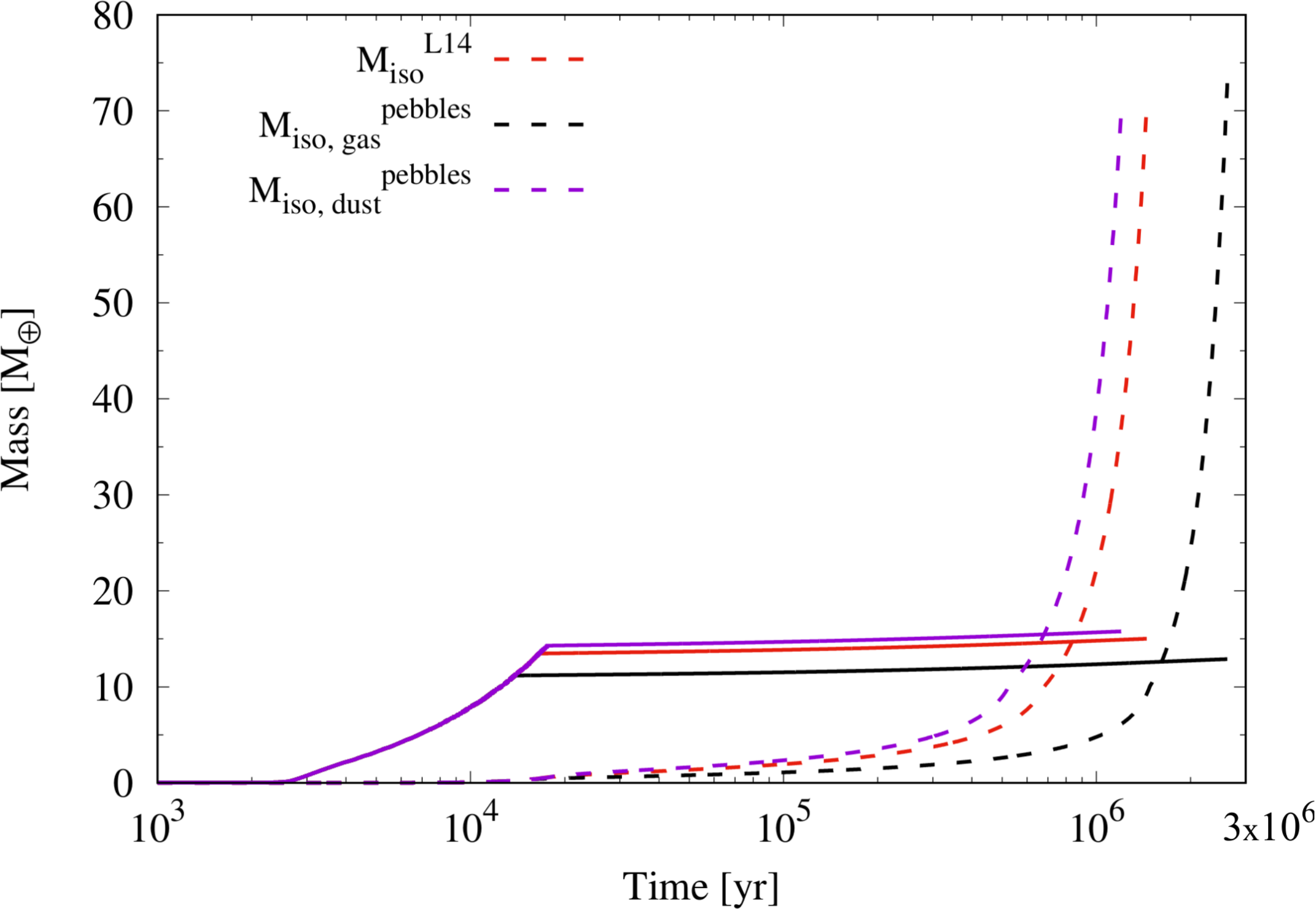} 
  \caption{Time evolution of the mass of the core and the envelope mass of the planet considering the pebble isolation masses given by \citet{Ataiee2018} comparing to the fiducial case where we adopted the pebble isolation masses from \citet{Lambrechts+2014}. $\text{M}_{\text{iso, gas}}^{\text{pebbles}}$ represents the pebble isolation mass estimated from pure gas hydrodynamical simulations, while $\text{M}_{\text{iso, dust}}^{\text{pebbles}}$ take into account also the dust filtration across the partial gap due to dust diffusion.}
  \label{fig1-sec3-3}
\end{figure}

In Fig.~\ref{fig1-sec3-3}, we plot the time evolution of the mass of the core and the mass of the envelope for the case where $\text{M}_{\text{iso, gas}}^{\text{pebbles}}$ and $\text{M}_{\text{iso, dust}}^{\text{pebbles}}$ are considered, comparing to the fiducial simulation. We found that $\text{M}_{\text{iso, gas}}^{\text{pebbles}}$ is $\sim 10~\text{M}_{\oplus}$, being a bit smaller compared to the one obtained in the fiducial simulation using the prescriptions from \citet{Lambrechts+2014}. However, we found that $\text{M}_{\text{iso, dust}}^{\text{pebbles}}$ gives us a pebble isolation mass larger, of $\sim 15~\text{M}_{\oplus}$, than the mentioned before. The  results show a good agreement despite that the pebble isolation mass proposed by \citet{Lambrechts+2014} does not take into account the dependence with the $\alpha$-viscosity parameter. Despite of the differences on the pebble isolation mass, the gas accretion rates onto the planet are similar, and formation time-scales are larger for smaller cores. We also note that all these studies performed hydrodynamical simulations for standard protoplanetary disks that do not present pressure bumps.  

\begin{figure}
  \centering
  \includegraphics[width= 0.5\textwidth]{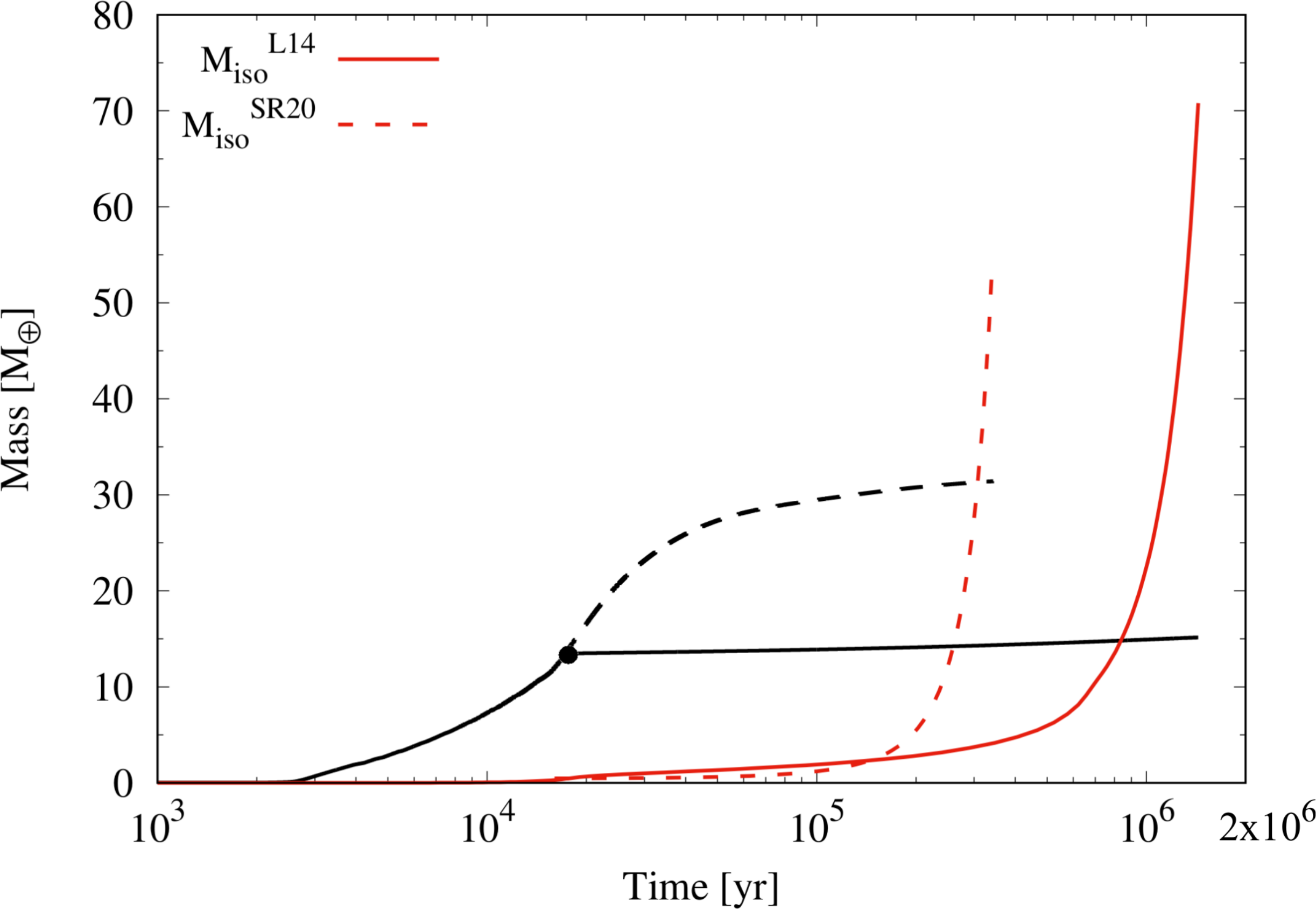} 
  \caption{Time evolution of the mass of the core ( the black lines) and the envelope (the red lines) of the planet for the fiducial simulation (the solid lines) and the case where the planet is allowed to accrete the pebbles at its co-orbital region after it reached the pebble isolation mass (the dashed lines).}
  \label{fig5-sec3-2}
\end{figure}

Recently, \citet{Sandor2020} investigate pebble accretion and pebble isolation mass in a generic pressure maximum of a protoplanetary disk performing two-fluid 2D hydrodynamical simulations of gas and dust, treating the dust as a pressureless fluid. They found that in this scenario, when the planet opens a partial gap and it halts the flux of incoming pebbles, there is a significant amount of pebbles (and gas) in the co-rotation region of the planet that can be accreted by it. In order to mimic this situation with our more simple disk model, we performed a simulation in which when the planet reaches the pebble isolation mass we set to zero the dust drift velocities, and hence we halted the incoming flux of pebbles. In this context, the planet can continue growing by accreting pebbles that remain in its feeding zone. In Fig.~\ref{fig5-sec3-2}, we can see that in this situation (the black dashed lines) the mass of the core continues growing until reach asymptotically $\sim 30~\text{M}_{\oplus}$, and the planet triggers the runaway gas accretion phase in only $\sim 3\times10^5$~yr. We remark that the growth of the core is qualitatively very similar to the results found by \citet{Sandor2020}, where for a planet growing at a pressure bump there is not a well defined pebble isolation mass. 

\section{Discussion}
\label{sec:sec4}

In this section, we discuss the implications of our model, and add some comparisons with previous work. 

\subsection{From dust to gas giants, including pebble and planetesimal accretion}
\label{sec:sec4-1}

In this work, we studied if a pressure maximum generated by a viscosity transition at the ice-line can act as an efficient location for dust trap, dust growth and planetesimal formation by streaming instability. We showed that indeed at the pressure maximum location dust is very efficiently accumulated. Since the Stokes numbers of pebbles is about $\mathrm{St} \sim 1$, these particles rapidly sediment to the disk's mid-plane, so that the ratio between the volumetric density of solids and gas exceeds unity, and streaming instability sets in. We also show that if an embryo is located at the planetesimal forming region, it can reach the pebble isolation mass in a time-scale of a few tens of thousand years.

The recent works of \citet{alibert2018} and \citet{Venturini-Helled2020} suggest that Jupiter could form by the hybrid accretion of pebbles and planetesimals. In this scenario, first a massive core is formed mainly by the accretion of pebbles. After the planet reaches the pebble isolation mass, the energy released by the accretion of planetesimals delays the time the planet can reach the gaseous runaway phase for a few million years. This scenario could naturally explain the time required for the formation of two reservoirs of small bodies in the early Solar System \citep{Kruijer2017}, and the atmospheric enrichment of Jupiter by \citet{Wahl2017}.

In this paper we additionally considered such a hybrid accretion scenario in a complete self-consistent manner.
We showed that the accretion of planetesimals after attaining the pebble isolation mass effectively delays the onset of the gaseous runaway phase during at least 1~Myr. As in \citet{alibert2018}, the energy released by the planetesimal accretion avoids the quick compression of the envelope. While in \citet{alibert2018} the planetesimal accretion rate is constant, in this work the planetesimal accretion rate decreases in time about two order of magnitude due to the planet's gravitational excitation onto the planetesimals, and due to the decrease in the planetesimal surface density by planet accretion. This is line with the results obtained by \citet{Venturini-Helled2020} who found a decreasing in the planetesimal accretion rates  for short orbital periods (for $R\lesssim3$~au). They also found that for larger distances the planet never empties its feeding zone, so the planetesimal accretion rate decreases much slower. Thus, the hybrid accretion scenario could delay the gas accretion for even longer times at a bit larger orbital periods. In any case, the qualitative picture for the formation of a giant planet is the same as in \citet{alibert2018} and here: a massive core is formed mainly by the accretion of pebbles and the accretion of planetesimals delays the planet to reach the gaseous runaway phase. Moreover, the giant planet formation picture in this hybrid scenario is qualitatively similar to the one found by \citet{P96}, with the difference that here the phase 1 (the first phase when a massive core is formed) is characterized mainly by the accretion of pebbles until the planet reaches the pebble isolation mass. 

Finally, we showed that the pressure maximum acts as a planet migration trap if type I migration recipes for isothermal disks are used. Despite of the fact that we are using a vertical isothermal disk, and that the temperature profile does not evolve in time, we also tested the previous result using planet migration recipes derived for non-isothermal disks. We found that using the migration recipes from \citet{jm2017}, the pressure bump continues working as a planet migration trap. However, if we adopt the migration recipes from \citet{paardekooper.etal2011}, the migration trap is broken when the mass of the planet is of about several tens of Earth masses. Thus, pressure bumps are not only preferential locations for planet growth, they can also act as planet migration traps helping giant planets to survive at moderate-large distances from the central star.

\subsection{Gas accretion onto the planet and the pebble isolation mass}
\label{sec:sec4-2}

In the fiducial simulations, we showed that if the solid accretion is halted after the planet reaches the pebble isolation mass at $\sim 13.5~\text{M}_{\oplus}$, the gas accretion runaway phase is triggered in only a few hundred thousands years. However, when the hybrid accretion of pebbles and planetesimals is considered, the heat released by the planetesimals delays the onset of the runaway of gas by $\sim 1.2$~Myr. We also studied the dependence of our results with the pebble isolation mass. \citet{Ataiee2018} studied the dependence of the pebble isolation mass with the alpha-viscosity parameter through hydrodynamical simulations, analyzing also how dust diffusivity can increase the pebble isolation mass found by pure gas hydro-simulations. We found that the pebble isolation mass derived for pure gas hydro-simulations is a bit lower than the computed in the fiducial simulations given by \citet{Lambrechts+2014}. As a consequence, the onset of the gas runaway phase is delayed in about a factor two. On the other hand, the pebble isolation mass computed taking into account the dust diffusion is larger than the fiducial case and the growth of the planet is very similar. Finally, \citet{Sandor2020} recently showed through hydrodynamical simulations of gas and dust, that in a pressure bump, when the planet opens a partial gap, there is still a significant amount of gas and pebbles in the planet's corotation region. They found that for Stokes numbers $\mathrm{St}\gtrsim 0.1$, the planet can accrete efficiently at least a part of this available material and that the pebbles isolation mass might not be a well defined quantity, due to the fact that the planet can grow asymptotically. Thus, we also mimic this scenario where the planet is allowed to accrete the pebbles that remain in the corotation region when it reaches the pebble isolation mass. We found a qualitatively similar results compare to \citet{Sandor2020}, and the planet core grows asymptotically to $\sim 30~\text{M}_{\oplus}$. In this case, because of the large mass of the core, the planet triggers the gaseous runaway phase in only a few hundred thousand years.

We note that in all our simulations the accreted solids reach the planet's core and do not pollute the H-He envelope. However, the gas accretion by the planet from the disk can be significantly modified if envelope enrichment by heavy elements is considered. \citet{Venturini+2016} showed that if the sublimated water coming from the icy pebbles or planetesimals is uniformly mixed along the envelope, the mean molecular weight of the envelope rises, leading to a more efficient envelope contraction that enhances gas accretion. In addition, \citet{Brouwers2018} and \citet{Brouwers2020} found similar results if the sublimated silicate by the accretion of dry-pebbles is well mixed at the bottom of the envelope. However, in this last case \citet{Bodenheimer2018} found that if a compositional gradient is formed between the fully saturated silicate vapor inner envelope and the non-polluted outer H-He envelope, the gas accretion by the planet from the disk decreases. Despite the fact that the incorporation of envelope enrichment is well beyond the scope of this paper, we plan to include this complex phenomenon in future works. 

\subsection{Transition viscosities as rings in protoplanetary disks}
\label{sec:sec4-3}

\begin{figure}
  \centering
  \includegraphics[width= 0.475\textwidth]{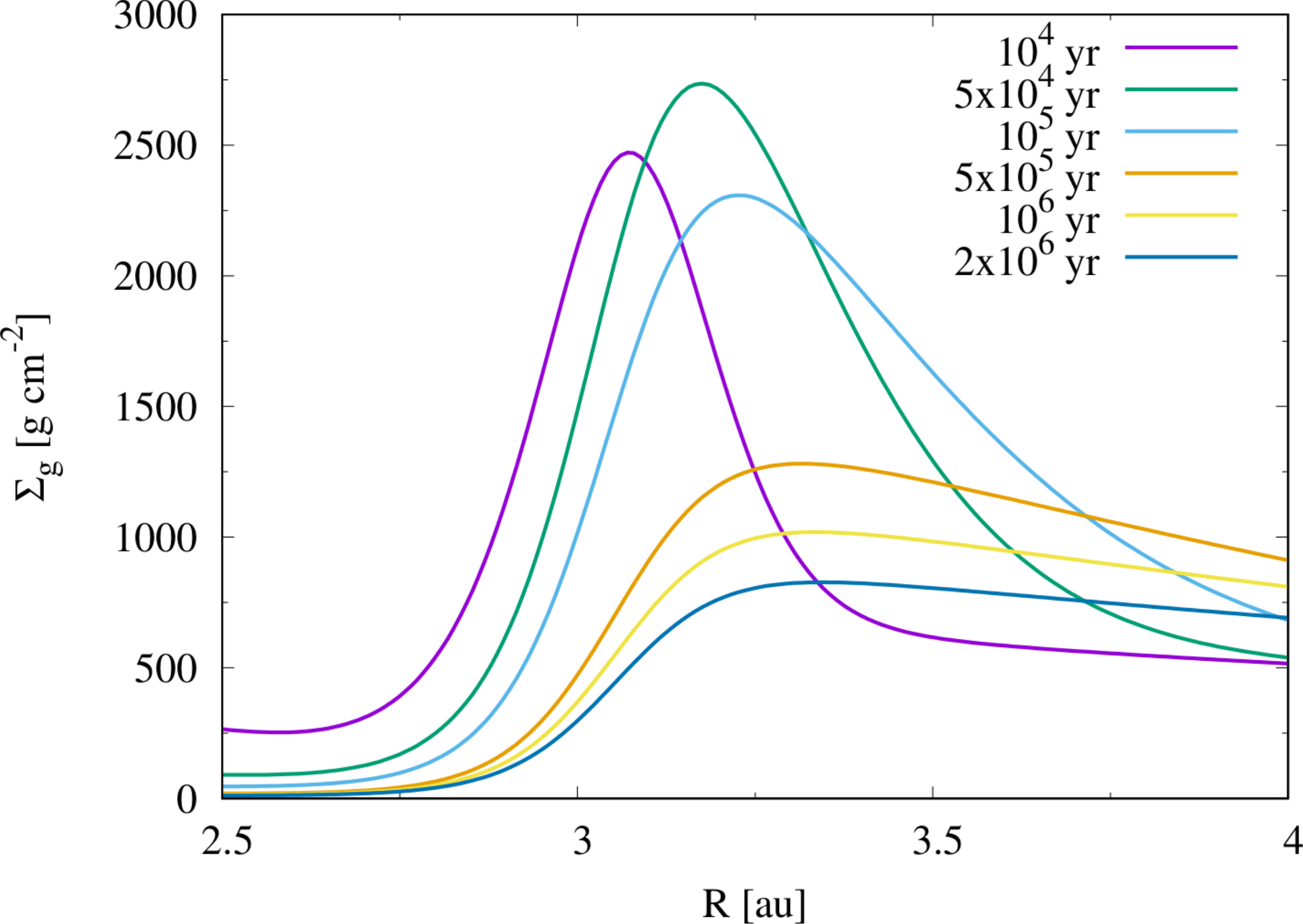} 
  \caption{Zoom around the viscosity-transition location of the time evolution of the radial profiles of the gas surface density for fiducial simulation.}
  \label{fig1-sec4-3}
\end{figure}

\begin{figure}
 \centering
 \includegraphics[width= 0.475\textwidth]{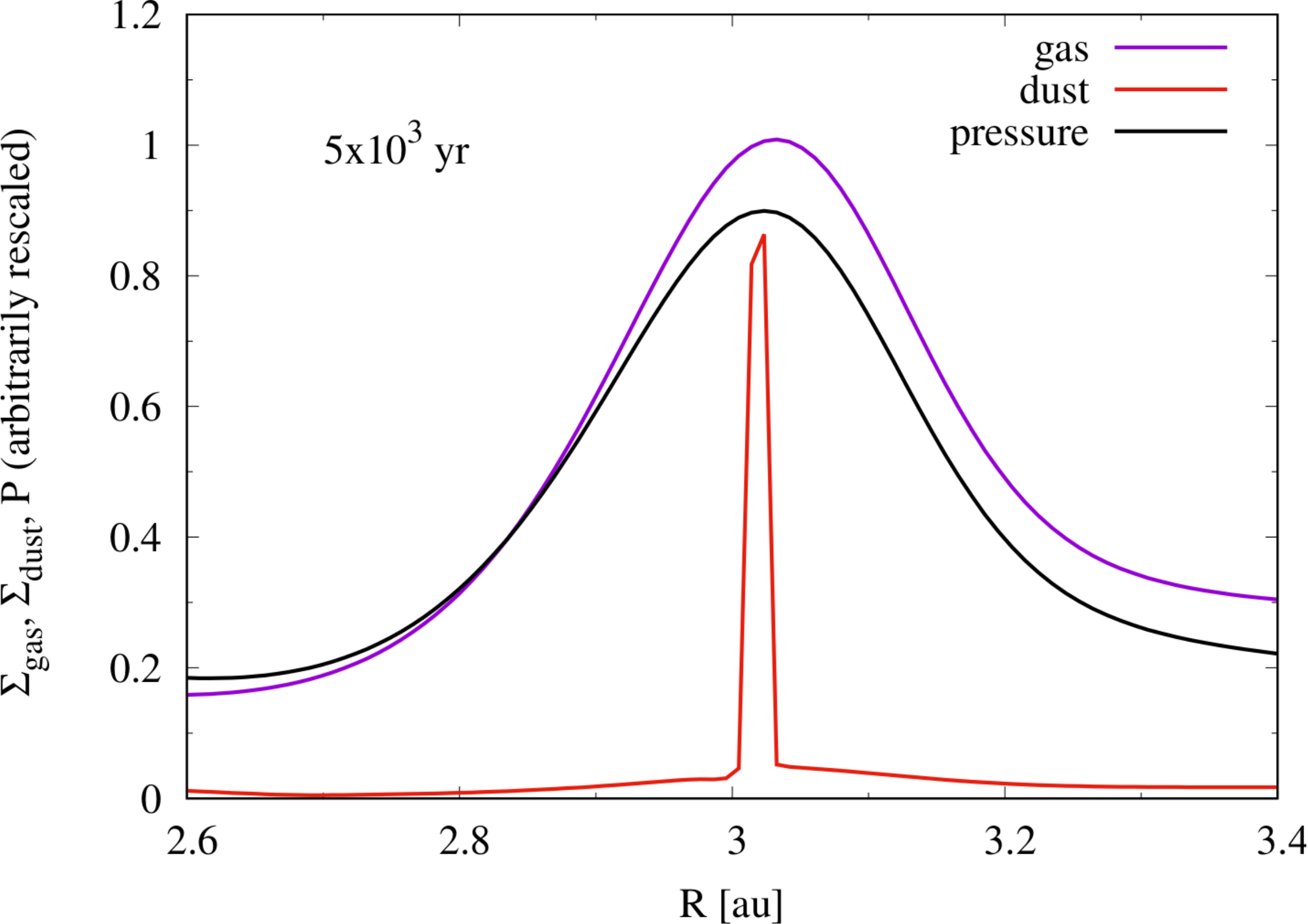} \\
 \includegraphics[width= 0.475\textwidth]{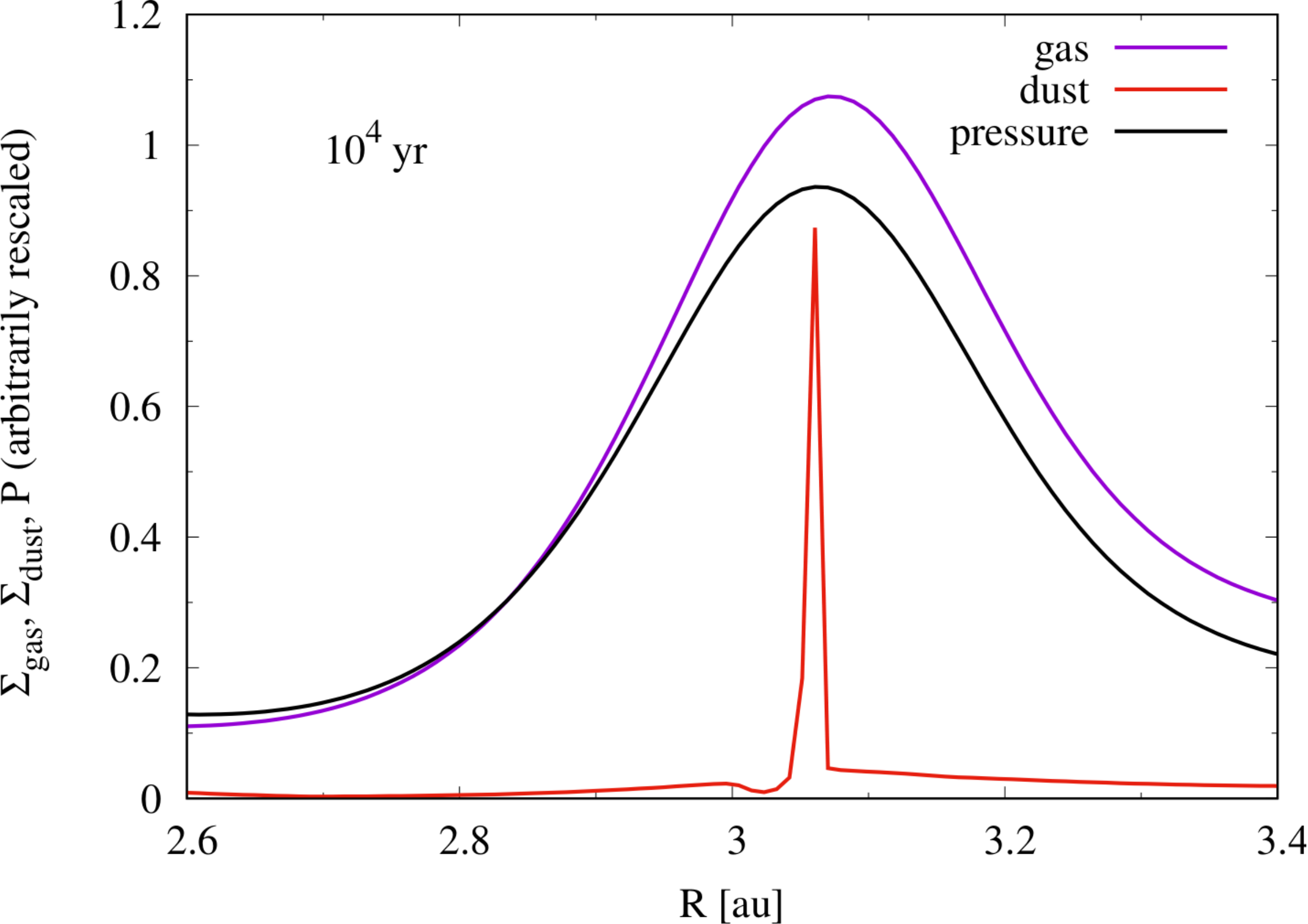} \\
 \includegraphics[width= 0.475\textwidth]{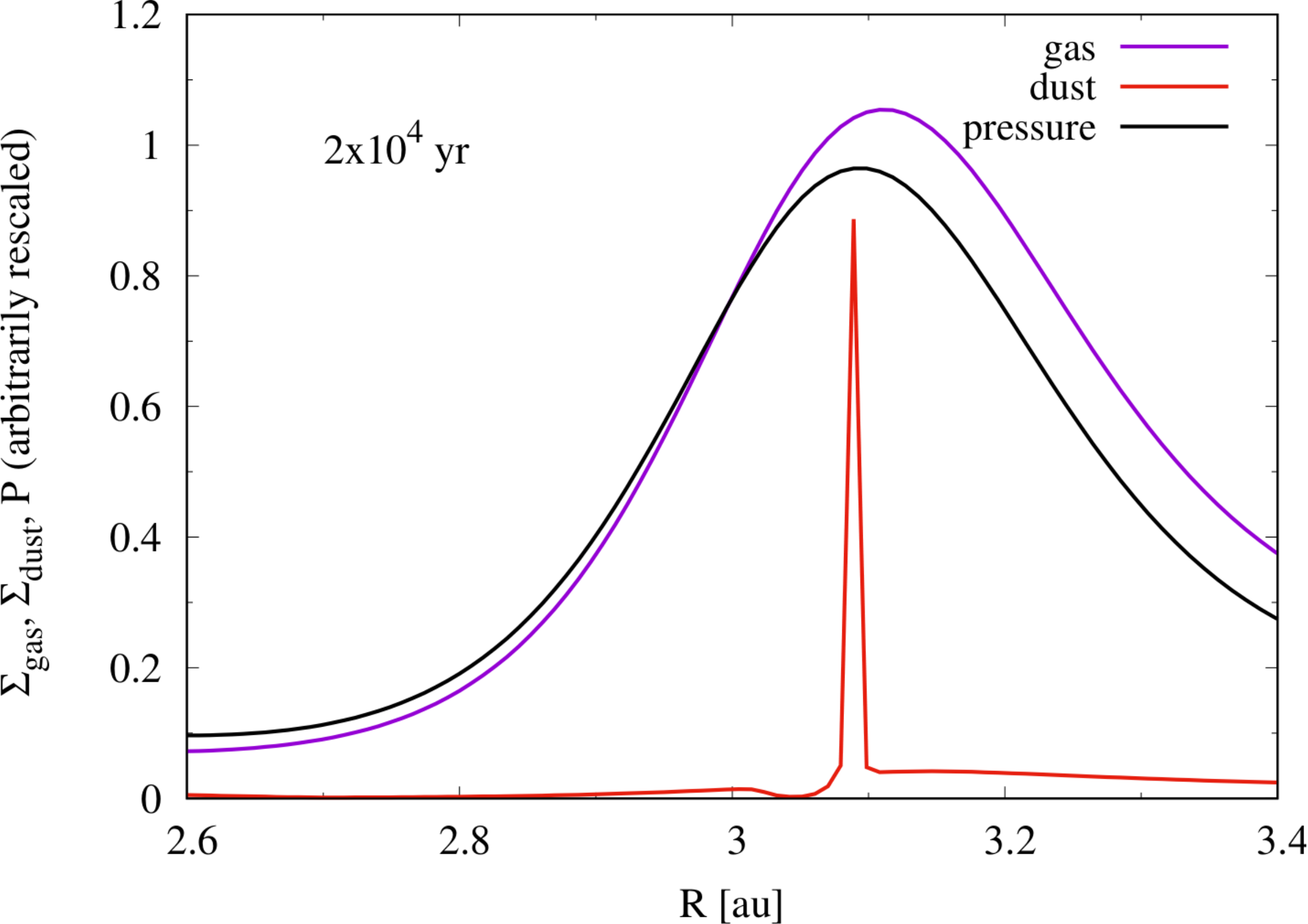} \\
  \caption{Radial profiles of the gas surface density, dust surface density, and the pressure at three different times until the planet reaches the pebble isolation mass. Gas and pressure maxima slightly move outwards as the disk evolves. The vertical scale of each profile was re-scaled for illustrative purposes as in \citet{Morbidelli2020}.}
  \label{fig2-sec4-3}
\end{figure}

Recently, \citet{Morbidelli2020} studied semi-analytically the formation of massive cores by pebble accretion in pressure bumps with gaussian profiles \citep{Dullemond2018}. He found that DSHARP (Disk Substructures at High Angular Resolution) rings are too far from the star to allow the formation of massive planets within the disk’s lifetime. However, a similar ring re-scaled to 5~au could lead to the formation of a massive core in less than 0.5~Myr, if pebble accretion results efficient. It is interesting for us to highlight that the transition viscosity adopted in this work also generates pseudo-gaussian profiles. This can be seen in Fig~\ref{fig1-sec4-3}, where we plot a zoom around the viscosity-transition region of the surface density of gas to show this. Thus, our study has some similarities to the one developed by \citet{Morbidelli2020}. To emphasize this, Fig.~\ref{fig2-sec4-3} shows three snapshots in time of the gas surface density, dust surface density, and pressure of the disk \citep[arbitrary re-scaled as in][]{Morbidelli2020} for the fiducial simulation until the planet reached the pebble isolation mass. We can see that the scenario is very similar to the one studied by \citet{Morbidelli2020}. Moreover, \citet{Morbidelli2020} also found that planet Type-I migration is stopped in the ring, but not necessarily at its center. In fact, this is the same result that was found by \citet{Guilera&Sandor2017} but using the isothermal type I recipes from \citet{Tanaka+2002}. In this new work, we are also numerically corroborating the results found semi-analytically by \citet{Morbidelli2020}: the pressure bump also acts as a type I migration trap adopting recipes derived for non-isothermal disks. Thus, this work can be regarded as a completion of the work of \citet{Morbidelli2020} including disk evolution, dust growth and evolution, planetesimal formation by streaming instability, and calculating the formation of a giant planet by the hybrid accretion of pebbles and planetesimals and the surrounding gas. So, if rings like the ones observed by DSHARP could be generated by viscosity-transitions in the disk, our new model could allow us to study planet formation in such rings, too. We will focus on this in a future work as well as expand the model to non-isothermal disks. 

\subsection{Pressure bumps as cradles of planet formation}
\label{sec:sec4-4}

The primary goal of this work was to investigate the whole formation process of only one giant planet at a pressure maximum generated by a viscosity transition at the water ice-line. However, if the formed planet can leave the density/pressure maximum quickly, the pressure maximum could still efficiently accumulate the remaining dust drifting inwards, leading to the formation of a second planet. Moreover, if multiple seeds are formed during the streaming instability \citep[e.g][]{Liu2019}, it would also be important to study in detail which of them can grow up to the size of a gas giant. 

In addition, there are two important phenomena that could also limit the growth of the planet. The first one is the longevity of the pressure maximum, i.e. how long the conditions are fulfilled to sustain a density/pressure maximum. This will determine whether cores massive enough to form gas giants can be born there, or only smaller mass icy/silicate planets with tenuous atmosphere. As an example, \citet{Guilera&Sandor2017} showed that a pressure maximum generated also by a viscosity transition at the outer edge of a dead zone disappears at some point due to the viscous disk evolution. 
The second one is the distance between the location of the pressure maximum and the zero torque location in the type I migration regime which, as showed by \citet{Guilera&Sandor2017}, is not zero in isothermal disks. If this distance is not small, the accretion of the solid material accumulated at the pressure maximum would not be efficient. This situation could also occur in non-isothermal disks because in general, these locations do not match.

We note that the location of the water ice-line could not be the only one where a pressure maximum can develop. Another option could be the place where, closer to the star, the disk becomes thermally ionized \citep[see][and references therein]{Armitage2013}. This happens at $T \sim 900$ K, when alkali metals are partially ionized. At this location the conductivity of the plasma is also changed that may result in mismatching accretion rates leading again to the formation of a density/pressure maximum. Accumulation of solid material can be expected here too. However, very close to the central star, where such temperatures can be found, the in situ formation of gas giants can be inhibited due to the small aspect ratio of the disc \citep[][]{Venturini2020b}. This leads mainly to the formation of Mini-Neptunes or at most, gas dwarf planets.

Furthermore, it is generally assumed that the ring structures recently observed with ALMA in protoplanetary disks \citep{DSHARP1} are formed at pressure maxima in gas. A recent work of \citet{DullemondPenzlin2018} has revealed by simple linear stability analysis that if dust enhancements are suppressing MRI turbulence, ring-like instabilities develop (we note here that regarding the MRI suppression, very similar physics works, as in the case of water ice-line, see more details in the Introduction.) If due to a small perturbation dust over density appears, the MRI is reduced in that place leading to a higher value of the gas surface density that collects more dust suppressing even more the MRI. This is a positive feedback mechanism that triggers a density/pressure maximum in gas able to collect dust favoring planet formation.
In addition, MHD simulations also showed that multiple pressure maxima, associated to instabilities driven by MHD winds,  can be developed along the disk, and that dust can be efficiently accumulated there forming dusty rings \citep[e.g.][]{Riols2019, Riols2020}. 

Finally, a pressure bump appears beyond the orbit of a protoplanet that reached the pebble isolation mass. This can trigger a chain of planet formation and play an important role in the formation of planetary systems.

If all possible places/mechanisms of density/pressure maxima are able to form planets, a complete planetary system might be born in a single protoplanetary disk. Our approach could be extended to study planet formation in the above mentioned locations of a protoplanetary disk and this is the goal of future works.

\section{Conclusions}
\label{sec:sec4}

In \citet{Guilera&Sandor2017}, we studied the formation of giant planets at pressure maxima generated at the edges of a dead zone due to a viscosity transition. In that work, we considered different homogeneous size populations of planetesimals (of 0.1~km, 1~km, 10~km, and 100~km of radii) and pebbles (of 1~cm radii) along the disk. We found that pressure maxima act as planetesimals and pebble traps and also as a planet migration trap. This situation favored the formation of giant planets at the location of the pressure maxima. 

In this new work, we extend our previous one including: 
\begin{itemize}
\item a model of dust growth and dust evolution including back reaction based in the works of \citet{Birnstiel+2011,Birnstiel+2012} and \citet{Drazkowska+2016};
\item the formation of planetesimals by streaming instability based in the work of \citet{Drazkowska+2016}, with the corresponding appearance of a lunar-mass embryo by mass conservation; 
\item the growth of the planet by the hybrid accretion of pebbles and planetesimals and the surrounding gas. 
\end{itemize}

With our new model we studied the formation of a giant planet at a pressure maximum generated by a viscosity transition at the water ice-line by the hybrid accretion of pebbles and planetesimals. Our main results are:

\begin{itemize}
\item dust is very efficiently collected at the pressure maximum, triggering the quick and efficient formation of planetesimals by streaming instability;
\item the planet reaches the pebble isolation mass in a time-scale of only $10^4$~yr, mainly by the accretion of pebbles;
\item the energy released by the accretion of planetesimals delays the onset of the gaseous runaway in at least 1~Myr;
\item the pressure maximum acts as a type I migration trap during most of the planet growth. \end{itemize}

A similar model including dust growth, dust evolution, planetesimal formation and planet formation was recently presented by \citet{Voelkel2020}. However, some differences exist between both models: \citet{Voelkel2020} did not include the back reaction to calculate the drift velocities of the dust population; their planetesimal formation model is based on the pebble flux \citep{Lenz2019} and not on the streaming instability mechanism; and the growth of the planet cores is due only to the accretion of planetesimals. Thus to our knowledge this is the first work considering self-consistently, dust growth and evolution, planetesimal formation, and the growth of the planet core by the hybrid accretion of pebbles and planetesimals. We plan to expand our model in future works to study the formation of multi-planetary systems including the full picture of planet formation, beginning from a disk of gas and dust and incorporating the most relevant physical phenomena involved in this process. 

Our results also highlight the importance of inhomogeneities in disk properties as catalysts of giant planet formation. The presence of a primordial giant in the disk, or simply its trace towards the central star, could create further inhomogeneities that trigger subsequent dust accumulation and planet formation.

\begin{acknowledgements}
The authors thank Joanna Dr\c{a}\.{z}kowska for useful discussions about the dust model details. The authors also thank Marc Brouwers, the referee of the work, for his review which helped us to improve and clarify the manuscript. OMG is partially support by the PICT 2018-0934 from ANPCyT, Argentina. OMG and M3B are partially supported by the PICT 2016-0053 from ANPCyT, Argentina. OMG acknowledges financial support from ISSI Bern in the  framework of the Visiting Scientist Program. OMG and MPR acknowledge financial support from the Iniciativa Cient\'{\i}fica Milenio (ICM) via the N\'ucleo Milenio de Formaci\'on Planetaria Grant. MPR also  acknowledges financial support provided by FONDECYT grant 3190336 and from CONICYT project Basal AFB-170002. ZsS thanks the support of the Hungarian National Research, Development and Innovation Office (NKFIH), under the grant K-119993. OMG acknowledges the hosting by IA-PUC as an invited researcher. OMG and JV are thankful to discussions held during the ISSI Team ''Ice giants: formation, internal structure and link to exoplanets''.  
\end{acknowledgements}

\bibliographystyle{aa} 
\bibliography{biblio} 


\appendix

\section{Validation of our dust growth, dust evolution and planetesimal formation model}
\label{apex-1}

\begin{figure}
  \centering
  \includegraphics[width= 0.475\textwidth]{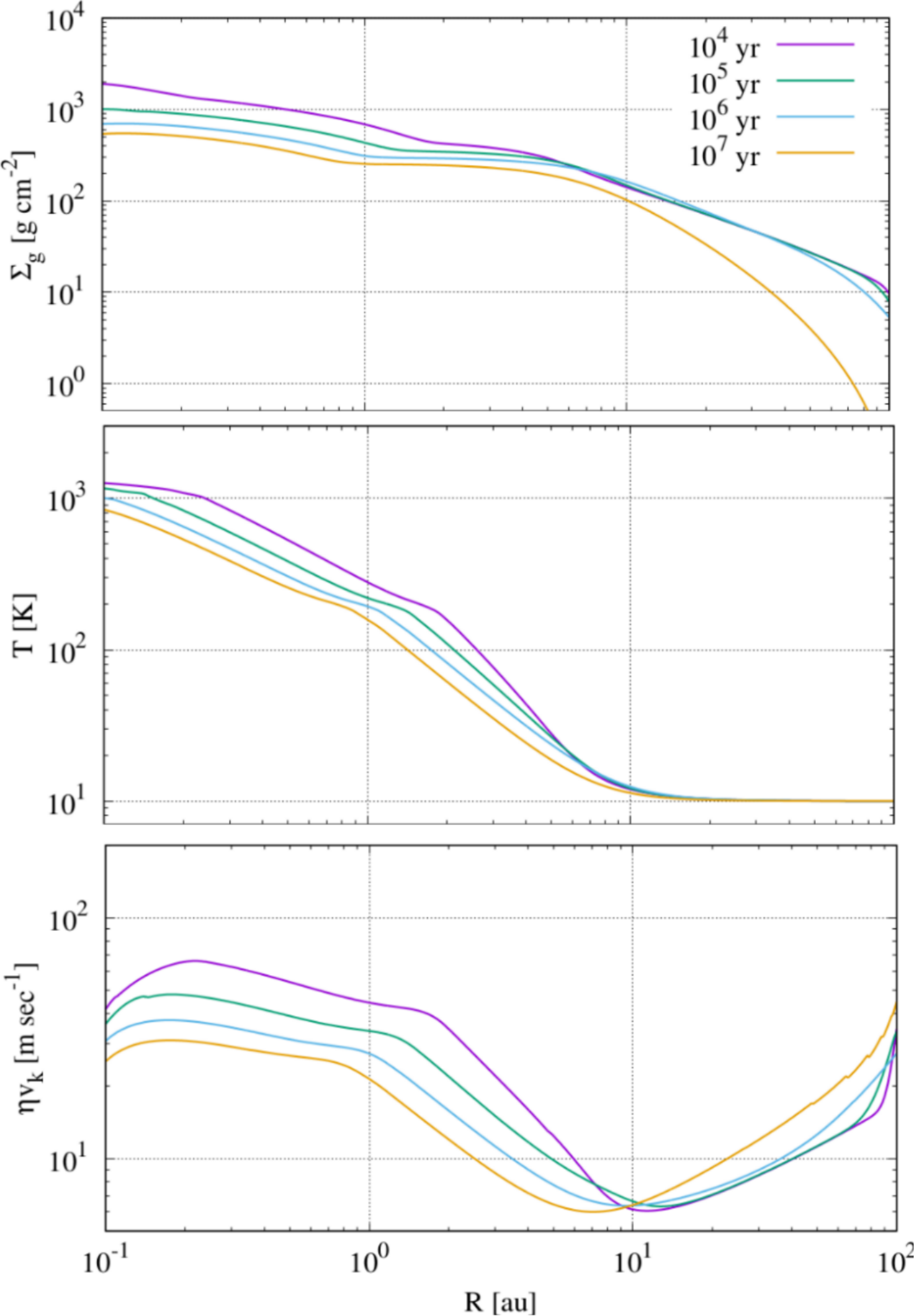} 
  \caption{Time evolution of the radial profiles of the gas surface density (top), temperature (middle), and the headwing $\eta v_k$ (bottom) for a disk similar to D16.}
  \label{fig1-apex-1}
\end{figure}

In order to validate the dust growth and evolution model added in {\scriptsize PLANETALP}, we tested it against the fiducial case of D16. That model involves a non-isothermal disk with an initial mass of $0.1~\text{M}_{\odot}$ that evolves viscously and by photoevaporation \citep{Veras2004}. The gas disk has an initial shallow surface density, corresponding to roughly $\Sigma_\text{g} \propto r^{-1}$, and after 10~Myr the disk has a mass of $\sim 0.02~\text{M}_{\odot}$. To compare our model with the one of D16, we also considered a non-isothermal disk adopting the same physics as in D16. We first computed the vertical structure of the disk using the same disk parameters of their fiducial case. We employed {\scriptsize PLANETALP} considering that the disk is only heated by viscosity, adopting a constant $\alpha$-viscosity parameter of $10^{-3}$ in all the disk (irradiation due to central star is neglected), and that heat is vertically transported only by radiation, in a similar way to the disk used in D16. In Fig.~\ref{fig1-apex-1}, we plot the time evolution of the radial profiles of the surface density of gas $\Sigma_{\text{g}}$, the temperature $\text{T}$, and the headwind $\eta v_k$. 
Our profiles and the ones presented in Fig.~2 of D16 look very similar,
despite some minor differences (mainly in the $\eta v_k$ profiles). 

\begin{figure}
  \centering
  \includegraphics[width= 0.475\textwidth]{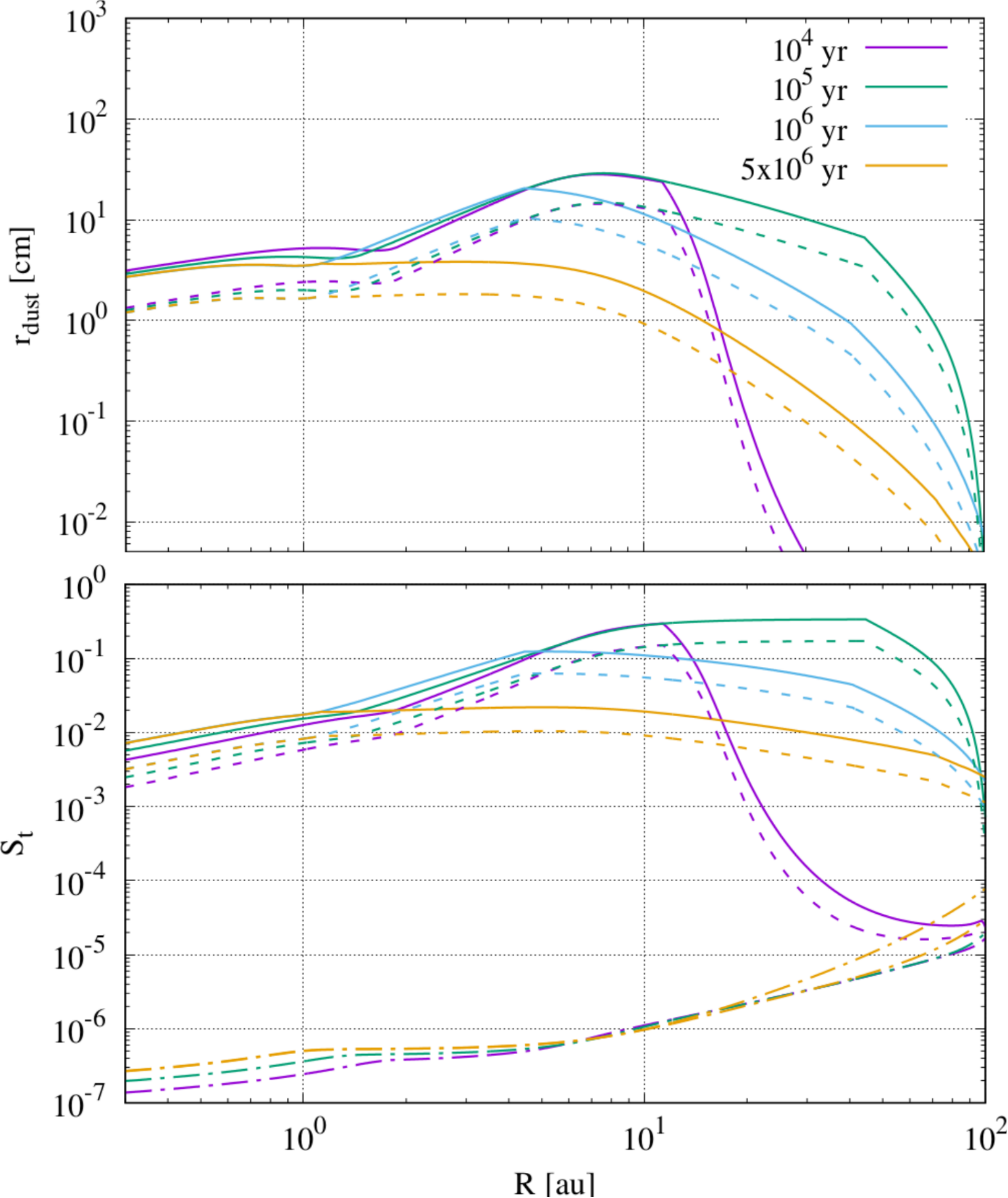} 
  \caption{Time evolution of the radial profiles of the dust size and Stokes number of the dust population. The solid lines represent the maximum values of the dust size and Stokes number. The dashed lines represent the corresponding weighted average of such quantities. The dashed-dot lines correspond to the minimum values of the Stokes numbers corresponding to the minimum value of the dust size of 1~$\mu$m.}
  \label{fig2-apex-1}
\end{figure}

In Fig.~\ref{fig2-apex-1}, we plot the time evolution of the radial profiles of the dust sizes and Stokes numbers. We note here that as in D16, we consider a threshold fragmentation velocity for the dust of 10 m/s along the disk. The solid lines represents the maximum size and Stokes number that the dust distribution reaches. For the case of the dust sizes, the maximum values and the locations of such maxima as the disk evolves are very similar to the ones presented in Fig.~3 in D16. We also show the weighted average of the dust size and the Stokes number (the dashed lines) and the corresponding minimum values (the dashed-dot lines). For the dust size, the minimum value is always $1~\mu$m along the disk. Since the dust mass distribution assigns most of the mass to the largest particles, the weighted average of the dust size and the Stokes numbers are close to the maximum values.

\begin{figure}
  \centering
  \includegraphics[width= 0.475\textwidth]{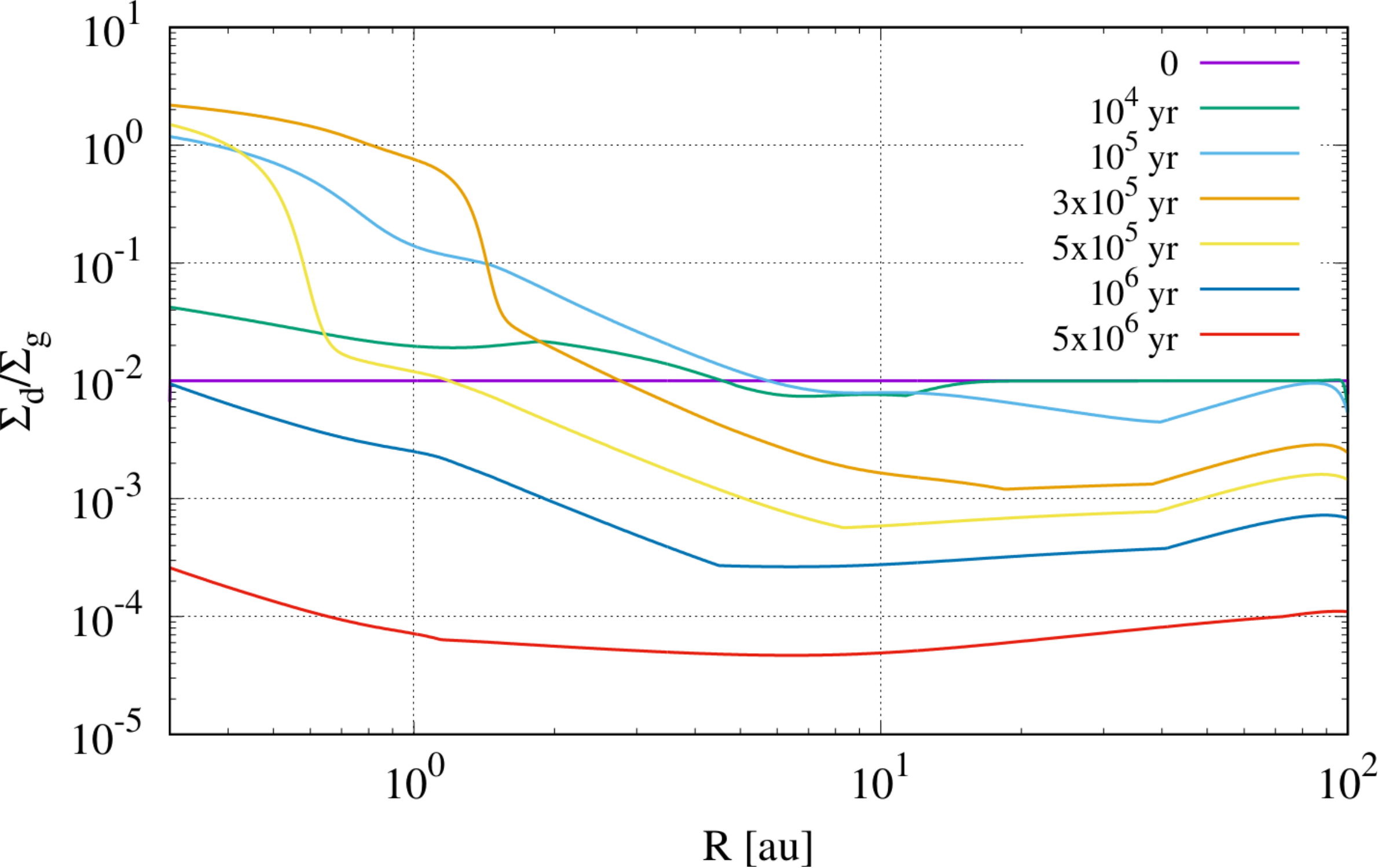} 
  \caption{Time evolution of the radial profiles of the dust-to-gas ratio. As in D16, the initial dust-to-gas ratio is set to 0.01 between 0.3~au and 100~au.}
  \label{fig3-apex-1}
\end{figure}

As in Fig. 4 of D16, in Fig.~\ref{fig3-apex-1} we present the time evolution of the radial profiles of the dust-to-gas ratio, initializing the dust-to-gas ratio equal to 0.01 between 0.3~au and 100~au. As time progresses, the dust accumulates in the inner part of the disk. This happens because the raise in the dust-to-gas ratio decreases the dust drift velocity. Although our radial profile at $3\times10^5$~yr is very similar to the one presented in Fig. 4 of D16, the profile at $10^6$~yr is quite different. In fact, our profile at $5\times10^5$~yr is very similar to the one of D16 at $10^6$~yr. We attribute this discrepancy to the fact that the initial conditions are not exactly the same for both models (the exact initial gas surface density profile is not specified in D16), and that the dust pile-up is strongly dependent on the temperature profiles (Joanna Dr\c{a}\.{z}kowska personal communication, also noticed by \citet{Birnstiel+2012}). Despite these differences, the time evolution of the radial profiles is qualitatively similar and present the same behavior. Moreover, the dust pile-up in the inner region allows the formation of planetesimals in the same region as in D16.

\begin{figure}
  \centering
  \includegraphics[width= 0.475\textwidth]{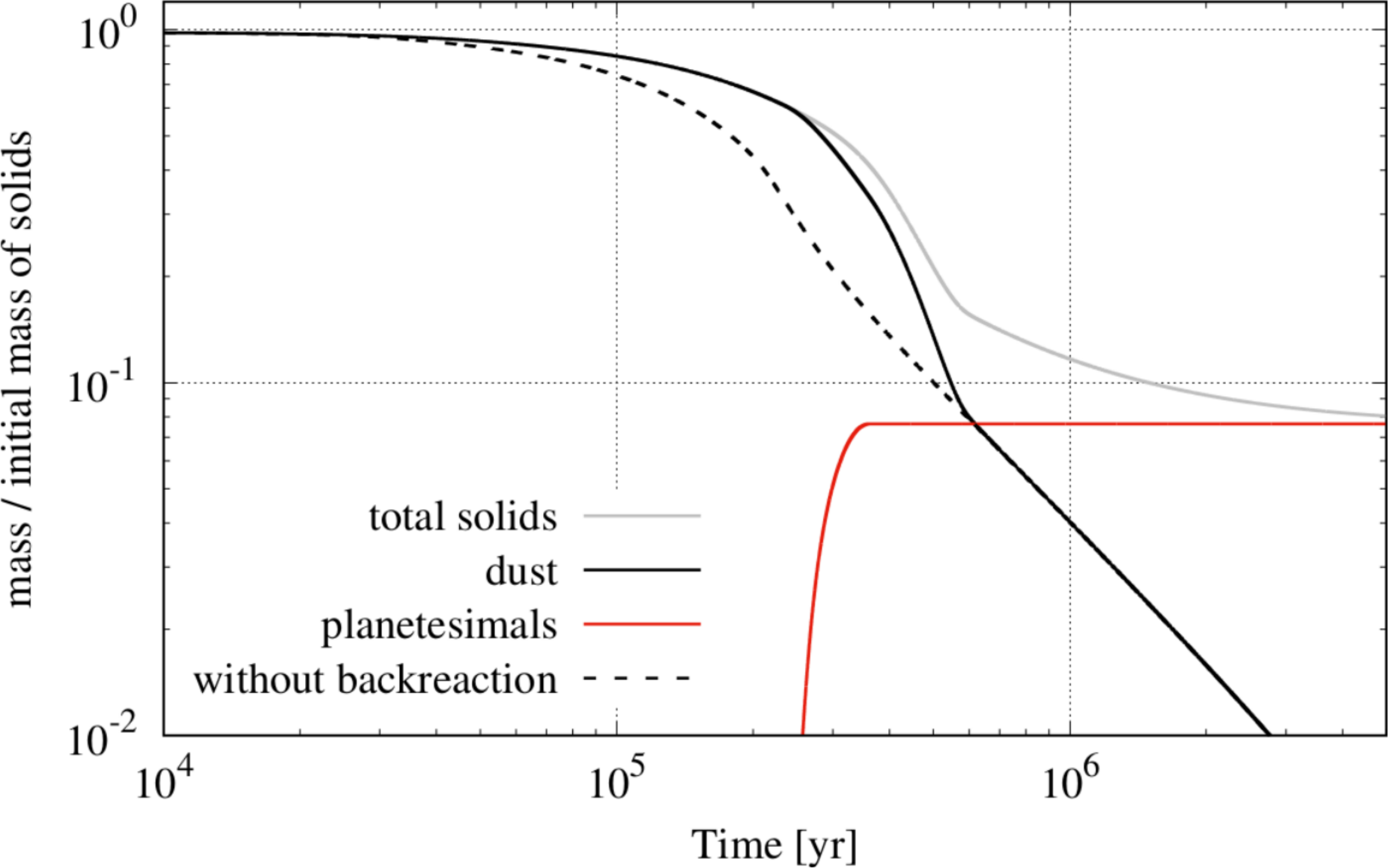} 
  \caption{Time evolution of the total mass of solids (grey line), the mass of dust (black line), and the mass of planetesimals (red line), all of them normalized to the initial mass of solid in the disk. The black dashed line represents the case where the backreaction is not considered. In this case, the dust pile-up in the inner region of the disk is less significant (due to the fact that dust drift velocities are higher) and the condition for planetesimal formation is not fulfilled. Thus, the dust mass represents the total mass.}
  \label{fig4-apex-1}
\end{figure}

Finally, in Fig.~\ref{fig4-apex-1} we plot the time evolution of the total mass of solids, the dust mass and the planetesimal mass. As in Fig.~5 of D16, if the back-reaction is not considered, the retention of dust is less efficient because drift velocities are higher. In contrast to D16, in our simulation only $\sim 8\%$ ($\sim 24~\text{M}_{\oplus}$) of the initial mass of solids is transformed into planetesimals. Again, despite the differences, the figures are qualitatively very similar and present the same behavior.  

\section{Different gas accretion models}
\label{apex-2}

The pioneering work of \citet{Ikoma2000} showed that if the solid accretion rate is halted at some point during the formation of the planet, gas accretion time-scales become shorter than $2.7 \times 10^5$~yr for cores above $10~\text{M}_{\oplus}$. They also calculated a semi-analytical prescription for gas accretion after solid accretion rate is halted, given by 
\begin{eqnarray}
\dot{\text{M}}_{\text{env}}= \dfrac{\text{M}_{\text{env}}^0}{\tau_{\text{g}}^{0}},
\label{eq1-apex-2}
\end{eqnarray}
$\text{M}_{\text{env}}^0$ being the envelope mass when solid accretion is halted, and $\tau_{\text{g}}^{0}$ is given by 
\begin{eqnarray}
\tau_{\text{g}}^{0}= 10^8 \left( \dfrac{\text{M}_{\text{C}}}{\text{M}_{\oplus}} \right)^{-2.5} \left( \dfrac{\kappa_{\text{gr}}}{1\text{cm}^2\text{g}^{-1}} \right).
\label{eq2-apex-2}
\end{eqnarray}
In this last expression, $\kappa_{\text{gr}}$ represents the grain opacity, and in order to fit their numerical results \citet{Ikoma2000} estimated that $\kappa_{\text{gr}}$ is given by \citep[see][]{Venturini2020b}  
\begin{eqnarray}
   \kappa_{\text{gr}}= 
   \begin{cases}
     0.22 \, f_{\text{gr}}~\text{cm}^2\text{g}^{-1} & \text{ if $170~\text{K} < \text{T}_{\text{sup}} < 1600~\text{K}$},  \\
     \\
     1 \, f_{\text{gr}}~\text{cm}^2\text{g}^{-1} & \text{ if $\text{T}_{\text{sup}} \le 170~\text{K}$},
   \end{cases} 
   \label{eq3-apex-2}
\end{eqnarray} 
$\text{T}_{\text{sup}}$ being the boundary temperature at the top of the envelope and $f_{\text{gr}}$ a reduction factor in the grain opacities respect to the values corresponding to the ISM. Despite the fact that several works propose a reduction of the grain opacity in the planet envelope due to grain growth and settling \citep[e.g][]{Movshovitz08,Mordasini04,Ormel14}, as in our model we use the grain opacities from \citet{Pollack1985}, which are representative for the ISM values, we set $f_{\text{gr}}= 1$, not considering a reduction in the grain opacities. Besides, as in the fiducial simulations the planet location remains always beyond the ice-line where $\text{T}_{\text{sup}}< 170$~K, we adopt $\kappa_{\text{gr}}= 1$. We note that a well known result is that as the grain opacities are reduced, the gas accretion onto the planet becomes more efficient \citep[e.g][]{Hubickyj2005}. Thus, a reduction in the grain opacities will reduce the formation timescales of the giant planets\footnote{As an example, in \citet{SanSebastian2019} we showed that the formation time of a giant planet growing at 5~au in a disk 10 times more massive than the MMSN \citep[Minimum Mass Solar Nebula,][]{Hayashi1981} by the accretion of 100~km planetesimals, is reduced in about 45\% if the grain opacities are reduced by a factor 50 in our model (see Fig.~13 of such work).}. 

Therefore, in order to compare the gas accretion by the planet, we compute a new set of simulations incorporating the gas accretion recipes from \citet{Ikoma2000}. We compute the planet growth using our fiducial model until the planet reaches the pebble isolation mass and then we compute the gas accretion following the prescriptions from \citet{Ikoma2000}. We perform two new simulations. In the first one, the accretion of solids is halted after the planet reaches the pebble isolation mass. For this case, we use Eq.~\ref{eq1-apex-2}, to compute the gas accretion by the planet after that time. In the second one, we consider the accretion of planetesimals after the planet reaches the pebble isolation mass. \citet{Ikoma2000} also showed that if the solid accretion rate is taken with a constant value of $10^{-6}~\text{M}_{\oplus}/\text{yr}$, the corresponding gas accretion time-scale, $\tau_{\text{g}}^\infty$, is about 6 times the gas accretion time-scale $\tau_{\text{g}}^{0}$. So, for simplicity, and despite the fact that the planetesimal accretion rate is not constant and in general greater than $10^{-6}~\text{M}_{\oplus}/\text{yr}$, we adopt $\tau_{\text{g}}^\infty$ to compute Eq.~\ref{eq1-apex-2} for this second simulation. 

\begin{figure}
  \centering
  \includegraphics[width= 0.475\textwidth]{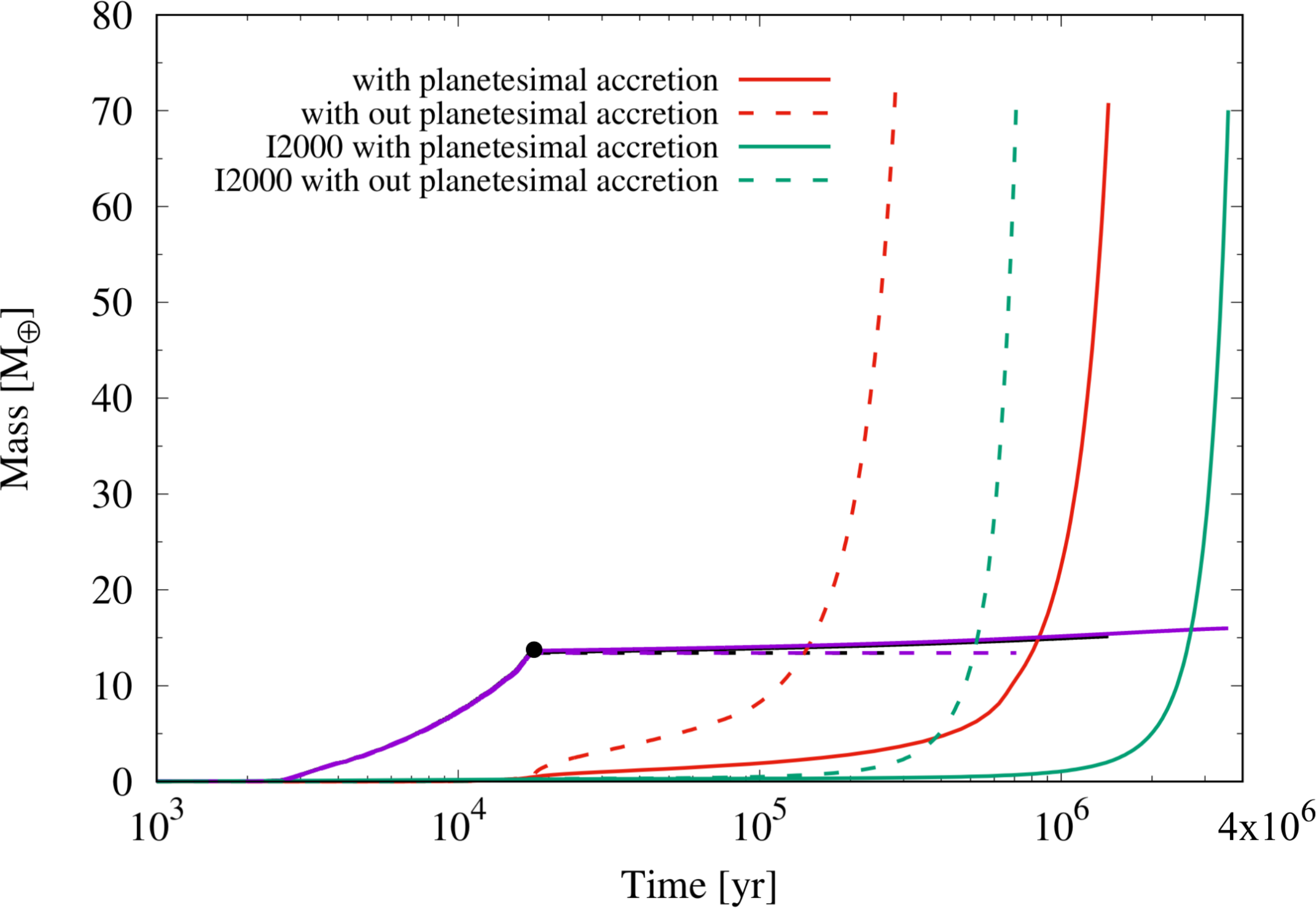} 
  \caption{Time evolution of the mass of the core and the envelope mass of the planet. The solid lines represent the models where the hybrid accretion of pebbles and planetesimals is considered, while dashed lines correspond to the cases where solid accretion is halted after the planet reaches the pebble isolation mass. The black (core) and red (envelope) lines correspond to the model where gas accretion is calculated with our full model \citep{Guilera2014}. The purple (core) and green (envelope) lines represent the models where the gas accretion rates from \citet{Ikoma2000} are used.}
  \label{fig1-apex-2}
\end{figure}

In Fig.~\ref{fig1-apex-2}, we plot the time evolution of the core mass and envelope mass comparing our fiducial simulations respect to the cases were gas accretion is computed following \citet{Ikoma2000} after the planet reaches the pebble isolation mass. As in Fig.~\ref{fig1-sec3-2}, the black and red curves represent the core mass and the envelope mass for the fiducial simulations, respectively. The purple and green lines represent the core mass and the envelope mass for the cases where we considered the gas accretion given by \citet{Ikoma2000}, respectively. In both sets of simulations, the dashed lines represent the case where solid accretion is halted after the planet reaches the pebble isolation mass, while the solid lines represent the case where planetesimal accretion is considered after that time. For the first case, we can see that the cross-over mass is reached at $\sim 5 \times 10^5$~yr (the dashed green line). We find that using our full model (the red dashed line), the planet accretes gas more efficiently and the runaway gas accretion phase is triggered a bit earlier, at $\sim 1.5 \times 10^5$~yr.  Still, the difference in the crossover time between the two approaches differs only by a factor of $\sim 3$. We find the same behavior when planetesimal accretion is considered after the planet reaches the pebble isolation mass. In this case, we find that with our model the planet reaches the cross-over mass at $\sim 9 \times 10^5$~yr (the red solid line), while employing the gas accretion recipes from \citet{Ikoma2000} the cross-over mass is reached at $\sim 2.7 \times 10^6$~yr. This implies also a delay in the formation timescale of a factor of $\sim 3$. Despite this difference, the results of considering the accretion of planetesimals is the same in both scenarios: to delay the planet to reach the gaseous runaway phase.

Summarizing, we found that solving the constitutive envelope equations with our model \citep{Guilera2010, Guilera2014} the cross-over mass is reached approximately three times earlier with respect to the case where the gas accretion prescription from \citet{Ikoma2000} is used. Despite the fact that the analytical recipes for the gas accretion from \citet{Ikoma2000} reproduce their numerical results, where also a Henyey method was applied to solve the constitutive envelope equations, some differences exist between both models. While in our model we use the equation of state (EoS) from \citet{Saumon1995}, \citet{Ikoma2000} used an ideal gas EoS considering hydrogen molecules and the ionization of hydrogen atoms. Other difference is related with the outer envelope boundary conditions. While in the work of \citet{Ikoma2000} these remain constant in time, in our model they evolve in time. The gas accretion rate is a very important quantity, because of it determines the time at which the gaseous runaway phase is triggered. Thus, a more detailed study about the differences in the results will be done in a future work. 

\end{document}